\documentclass[%
 superscriptaddress,
preprint,
 amsmath,amssymb,
 aps,longbibliography
]{revtex4-1}

\usepackage{graphicx}
\usepackage{dcolumn}
\usepackage{bm}
\usepackage{hyperref}
\usepackage[mathlines]{lineno}
\usepackage[bottom]{footmisc}
\usepackage{booktabs}
\usepackage[super]{nth}%

\begin{document}

\title{Physics-informed neural networks for inverse problems in nano-optics and metamaterials}

\author{Yuyao Chen}
\affiliation{Department of Electrical \& Computer Engineering and Photonics Center, Boston University, 8 Saint Mary's Street, Boston, MA 02215, USA.}
\author{Lu Lu}
\affiliation{Division of Applied Mathematics, Brown University, 170 Hope Street, Providence,  Rhode Island, 02912, USA.}
\author{George Em Karniadakis}
\affiliation{Division of Applied Mathematics, Brown University, 170 Hope Street, Providence,  Rhode Island, 02912, USA.}

\author{Luca  Dal Negro}
\email{dalnegro@bu.edu}
\affiliation{Department of Electrical \& Computer Engineering and Photonics Center, Boston University, 8 Saint Mary's Street, Boston, MA 02215, USA.}
\affiliation{Division of Applied Mathematics, Brown University, 170 Hope Street, Providence,  Rhode Island, 02912, USA.}
\affiliation{Division of Material Science and Engineering, Boston University, 15 Saint Mary's Street, Brookline, MA 02446, USA.}
\affiliation{Department of Physics, Boston University, 590 Commonwealth Avenue, Boston, MA 02215, USA.}

\begin{abstract}
In this paper we employ the emerging paradigm of physics-informed neural networks (PINNs) for the solution of representative inverse scattering problems in photonic metamaterials and nano-optics technologies. In particular, we successfully apply mesh-free PINNs to the difficult task of retrieving the effective permittivity parameters of a number of finite-size scattering systems that involve many interacting nanostructures as well as multi-component nanoparticles. Our methodology is fully validated by numerical simulations based on the Finite Element Method (FEM). The development of physics-informed deep learning techniques for inverse scattering can enable the design of novel functional nanostructures and significantly broaden the design space of metamaterials by naturally accounting for radiation and finite-size effects beyond the limitations of traditional effective medium theories.  
\end{abstract}

\maketitle
\section{Introduction}
For over two centuries, the science and engineering of electromagnetic waves in optical materials relied on either analytical solutions or numerical approximations of differential equations derived from physical models. While enormously successful, this approach fails to capture the multi-scale behavior of heterogeneous media whose structural complexity prevents the precise formulation and hence the solution of the high-frequency inverse scattering problem, which has numerous applications to optics, acoustics, geophysics, astronomy, medical imaging, microscopy, remote sensing, and nondestructive testing. This problem consists of determining the characteristics of an object from a \textit{limited set} of measured data on how it scatters incoming radiation; solving this may enable the predictive design of artificial optical materials, i.e. metamaterials and complex optical nanostructures, starting directly from desired optical functionalities within a prescribed frequency range. However, in the presence of strong multiple light scattering the inversion of the physics-driven differential models of light scattering in complex multi-particle geometries becomes an intrinsically ill-posed problem. 
Under these circumstances, traditional numerical techniques fail to predict desired systems' parameters. In addition, optical nonlinearities and unavoidable noise render object reconstruction from available data a \textit{computationally intractable problem}. 
These fundamental challenges motivated the  development of alternative and more powerful computational frameworks that leverage sophisticated optimization methods for the solution of inverse scattering problems deep in the multiple scattering regime \cite{kamilov2015learning,molesky2018inverse,liu2017seagle, kamilov2017plug, Pham2018Versatile, colton2018looking}. 
Very recently, the growing interest for machine learning algorithmic development in wave engineering led to successful applications to  inverse scattering problems in imaging and tomography \cite{sun2018efficient, sanghvi2019embedding}, demonstrating the potential to open new territories with respect to medical imaging, microscopy, and remote sensing technologies. 

Physics-informed neural networks (PINNs) \cite{Raissi,lu2019deepxde} is a general framework developed recently for solving both forward and inverse problems of partial differential equations. In this paper we propose and develop PINNs for the solution of different inverse scattering electromagnetic problems of direct interest to nano-optics and metamaterials technologies. In particular, we address the difficult problems of effective medium determination and parameter retrieval by applying the powerful PINNs framework to finite-size clusters of scattering nanocylinders arranged in periodic and aperiodic geometries. Moreover, using PINNs we demonstrate the ability to efficiently reconstruct the spatial distribution of the electric permittivity of unknown objects from synthetic scattering data. Finally, we apply our method to the determination of the optimal dielectric permittivity of optical cloaking layers beyond the quasi-static limit.  All our results are validated using numerical simulations based on the Finite Element Method (FEM) in two spatial dimensions.

Our work provides a novel and powerful framework for the inverse design of scattering nanostructures and photonic metamaterials based on the inversion of their scattered fields. Moreover, this ability establishes remote sensing functionalities in the optical regime whereby the unknown dielectric permittivity of complex optical nanostructures can be unambiguously retrieved from near-field optical imaging data.

\section{Physics-informed neural networks}

Our approach leverages the capabilities of deep neural networks as universal function approximators. However, differently from standard deep learning approaches, PINNs restrict the space of admissible solutions by enforcing the validity of partial differential equation (PDE) models governing the actual physics of the problem. 
This is achieved within relatively simple feed-forward neural network architectures leveraging automatic differentiation (AD) techniques readily available in the TensorFlow learning package \cite{Abadi}. Moreover, PINNs use only one training dataset to achieve the desired solutions, thus relaxing the burdens often imposed by the massive training datasets necessary in alternative, i.e. non physics-constrained, deep learning approaches. Moreover, PINNs do not require any data on the inverse parameters it predicts, and thus it belongs to unsupervised learning. These unique features of PINNs are greatly beneficial for the solution of inverse scattering problems either with measured field data or with synthetic ones generated by forward simulations.
Importantly, PINNs solve highly-nonlinear and dispersive inverse problems on the same footing as forward problems, by the simple addition of an extra loss term to the overall loss function for the minimization of the residuals of PDEs and their boundary conditions \cite{Raissi}. Therefore, PINNs are particularly
effective in solving {\em ill-posed} inverse problems, which are often intractable with available mathematical formulations.

\begin{figure}[h!]
\centering\includegraphics[width=10cm]{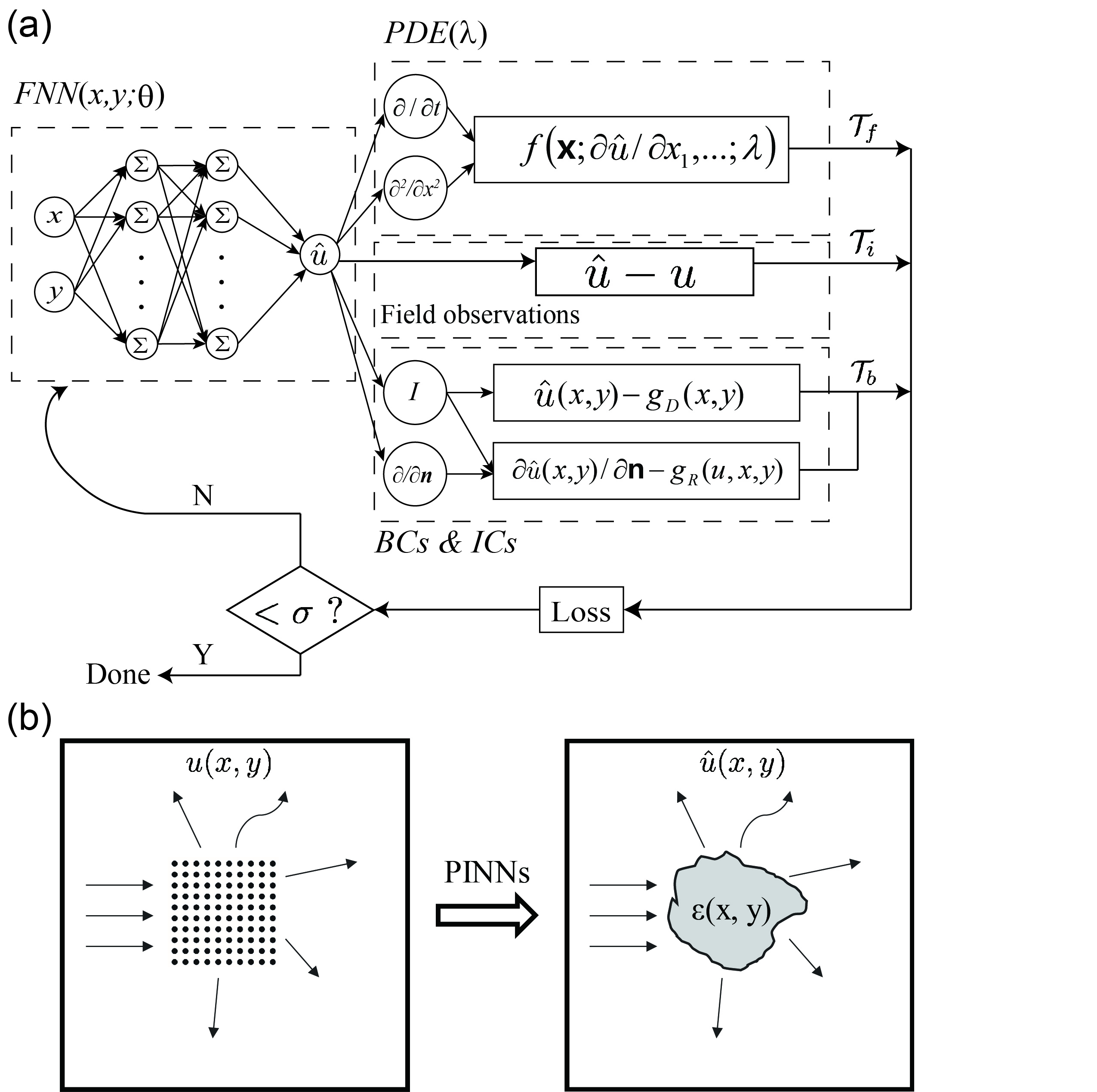}
\caption{(a) Schematic of a PINN for solving inverse problem in photonics based on partial differential equations. The left part neural network represents a surrogate model $u$ of the PDE solution. The right part shows the loss function to restrict $u$ to satisfy PDE in the domain $\Omega$, boundary conditions (BC) $\hat{u}(x,y)=g_{D}(x,y)$ on $\Gamma_{D} \subset \partial \Omega$, and $\frac{\partial \hat{u}}{\partial \mathbf{n}}(x, y)=g_{D}(u, x, y)$ on $\Gamma_{R} \subset \partial \Omega$. The initial condition (IC) is treated as a special type of boundary conditions. For inverse problem, we have also loss term from $\hat{u}-u$ on residual points. We optimize the neural network's weights and biases to obtain loss smaller than $\sigma$. (b) The process shows the process of PINNs to reconstruct the permittivity profile $\varepsilon$ from a known scattering field profile dataset.}\label{Fig1}
\end{figure}

In this section, we introduce the algorithm for solving inverse PDE models with PINNs. The main idea of PINNs is schematically illustrated in Fig. \ref{Fig1} (a). We first construct a neural network with the output $\hat{u}(\mathbf{x};\boldsymbol{\theta})$ as a surrogate of the PDE solution $u(\mathbf{x})$. Here, we employ the feed-forward neural network (FNN), which is relatively simple but sufficient for most PDE problems, and $\boldsymbol{\theta}$ is a vector containing all weights and biases in the neural network that needs to be trained. Then, the next key step is to constrain the neural network's output $\hat{u}$ to satisfy the PDE as well as data observations $u$. This is realized by constructing the loss function by considering terms corresponding to PDE, boundary conditions and initial conditions, and the field observations.
Specifically, we consider the following PDE problem with unknown parameter $\lambda$ for the solution $u(\mathbf{x})$ with $\mathbf{x} = {x_1,~\dots,~x_d}$ defined on a domain $\Omega \subset \mathbb{R}^d$:
\begin{equation}\label{eq:pde}
f\left( \mathbf{x}; \frac{\partial \hat{u}}{\partial x_1},  \dots, \frac{\partial \hat{u}}{\partial x_d}; \frac{\partial^2 \hat{u}}{\partial x_1\partial x_1},  \dots, \frac{\partial^2 \hat{u}}{\partial x_1\partial x_d};\dots; {\lambda} \right) = 0, \quad \mathbf{x} \in \Omega.
\end{equation}
We denote $\mathcal{T}_i \subset \Omega$ be the points where we have the PDE solution $u(\mathbf{x})$. The loss function is then defined as:
\begin{equation}\label{eq:loss}
\mathcal{L}(\boldsymbol{\theta},\lambda) = w_f \mathcal{L}_f(\boldsymbol{\theta},\lambda; \mathcal{T}_f) + w_i \mathcal{L}_i(\boldsymbol{\theta},\lambda; \mathcal{T}_i)+ w_b \mathcal{L}_b(\boldsymbol{\theta},\lambda; \mathcal{T}_b),
\end{equation}
where:
\begin{align}
\mathcal{L}_f(\boldsymbol{\theta},\lambda; \mathcal{T}_f) &= \frac{1}{|\mathcal{T}_f|} \sum_{\mathbf{x} \in \mathcal{T}_f} \left\Vert f\left( \mathbf{x}; \frac{\partial \hat{u}}{\partial x_1},  \dots, \frac{\partial \hat{u}}{\partial x_d}; \frac{\partial^2 \hat{u}}{\partial x_1\partial x_1}, \dots, \frac{\partial^2 \hat{u}}{\partial x_1\partial x_d};\dots;{\lambda} \right) \right\Vert_2^2, \\
\mathcal{L}_{i}(\boldsymbol{\theta},{\lambda}; \mathcal{T}_i) &= \frac{1}{|\mathcal{T}_i|} \sum_{\mathbf{x} \in \mathcal{T}_i} \| \hat{u}(\mathbf{x}) - u(\mathbf{x}) \|_2^2,\\\mathcal{L}_{b}\left(\boldsymbol{\theta},\lambda; \mathcal{T}_{b}\right)&=\frac{1}{\left|\mathcal{T}_{b}\right|} \sum_{\mathbf{x} \in \mathcal{T}_{b}}\|\mathcal{B}(\hat{u}, \mathbf{x})\|_{2}^{2},
\end{align}
and $w_f$, $w_b$, and $w_i$ are the weights. $\mathcal{T}_{f}$, $\mathcal{T}_{i}$, $\mathcal{T}_{b}$ denote the residual points from partial differential equations, training dataset, and ICs and BCs, respectively. Here, $\mathcal{T}_f \subset \Omega$ is a set of predefined points to measure the matching degree of the neural network output $\hat{u}$ to the PDE. $\mathcal{T}_f$ can be chosen as grid points or random points, see more details and discussion in reference \cite{lu2019deepxde}. In the last step, we optimize $\theta$ and $\lambda$ by minimizing the loss function $\mathcal{L}(\boldsymbol{\theta},\lambda)$.
Note that using PINNs the only difference between a forward and an inverse problem is the addition of an extra loss term $\mathcal{L}_{i}$ to Eq. \ref{eq:loss}, which comes at an insignificant computational cost. We use DeepXDE \cite{lu2019deepxde}, a user-friendly Python library, for the code implementation in this paper.

\section{PINNs for the homogenization of finite-size metamaterials}
Photonic metamaterials are artificial structures composed of designed subwavelength building blocks that can achieve unusual optical properties \cite{soukoulis2011past}. The current underpinning of metamaterials design relies on the effective medium theory (EMT) \cite{sheng2006introduction,choy2015effective,sihvola1999electromagnetic}, which is actively investigated in electromagnetic research from different perspectives such as the methods of coherent potential approximation (CPA) \cite{wu2006effective,zhang2013effective}, the multipole expansion method \cite{chremmos2015effective}, or the field averaging procedure \cite{gozhenko2013homogenization} are used to homogenize metamaterials properties. Here we propose to use PINNs for the homogenization of dielectric metamaterials formulated as an inverse medium problem for finite-size systems. Specifically, we set up large clusters composed of several hundreds dielectric nanocylinders arranged in periodic and aperiodic geometries and demonstrate, using PINNs, direct retrieval of the effective permittivity distribution that gives rise to the same scattered field as the original cluster. We refer to this approach as "inverse metamaterials design". 

In order to achieve our goals we constrained PINNs using the Helmholtz equation for weakly inhomogeneous two-dimensional (2D) media under TM polarization excitation:
\begin{equation}\label{inver homogenization}
    \nabla^{2}{E_z\left(x,y\right)}+{\varepsilon_{r}}\left(x,y\right)k_{0}^{2}E_{z}=0,
\end{equation}
where $E_z$ is the electric field $z$ component, $\varepsilon_{r} (x,y)$ is the space dependent relative permittivity, and $k_0=\frac{2\pi}{\lambda_0}$ is the wavenumber in free space. We first consider dielectric materials with real permittivity values and ignore the radiation losses in the effective medium, which will be addressed later in this section. However, as discussed at the end of the section, our method can be easily generalized to include lossy component materials as well. As schematically illustrated in Fig. 1(b), the problem at hand is the one to retrieve the function  $\varepsilon(x,y)$, which corresponds to the unknown parameters $\lambda$ in Fig. \ref{Fig1}(a), by training PINNs on synthetic data obtained via the forward solution $u(x,y)$ of the scattering problem using FEM. Validation of the permittivity spatial profile predicted by PINNs is obtained using the retrieved $\varepsilon(x,y)$ within a forward scattering FEM simulation. The resulting electric field distribution is then compared to the one of the original forward FEM simulation used to generate the training (synthetic) scattering data. The discrepancy between these two datasets is quantified by computing the $L^{2}$ error norm of the corresponding total field distributions. It is important to realize that in this section we train PINNs to predict the permittivity profile across the entire computational domain and therefore in this case we do not need to explicitly consider the loss term for the boundary conditions $\mathcal{T}_{b}$. 

\begin{figure}[h!]
\centering\includegraphics[width=10cm]{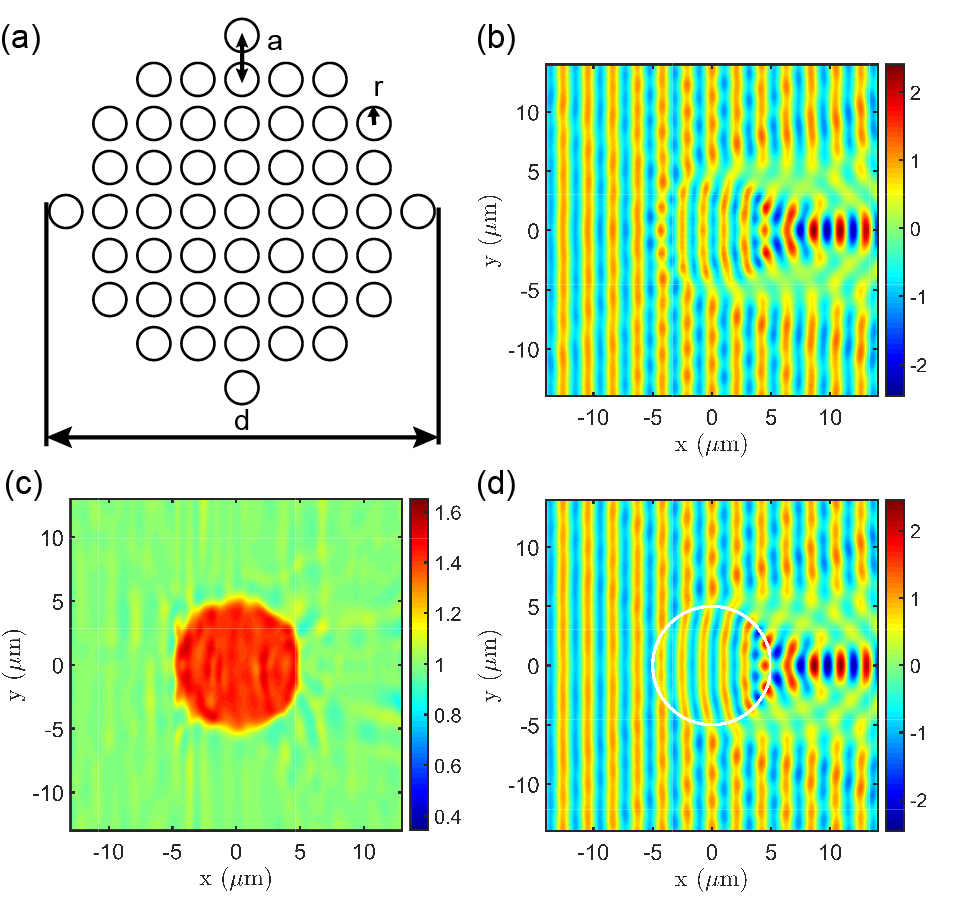}
\caption{(a) FEM simulation of electric field real part distribution profile for square lattice arranged scatters with $a=500nm$, $r=125nm$, $d=10{\mu}m$, $N=317$, $\epsilon=3$. The illumination plane wave wavelength is $\lambda=2.1{\mu}m$. (b) depicts the inverse train dataset for training PINNs to homogenize from panel (a). (c) shows the PINNs predicted permittivity profile by training with dataset shown in panel (b). (d) illustrates the FEM validation of electric field distribution when illuminating plane wave with wavelength $\lambda$ on the object with PINNs predicted permittivity profile. The $L^2$ error between the panel (b) and (d) is $2.82\%$.}\label{Fig2}
\end{figure}

We show in Fig. \ref{Fig2} (a) the representative schematics of a square lattice structure used for inverse metamaterial design, where $a$ denotes the center-to-center distance of the nanocylinders in the array, and $r$ is radius of each nanocylinder. The array pattern is truncated using a circular window with diameter $d$. Fig. \ref{Fig2} (b) illustrates the real part of the computed electric field distribution $E_z$ using FEM for a square lattice with parameters: $a=500nm$, $r=125nm$, $d=10{\mu}m$, number of particles $N=317$, and permittivity of nanocylinder $\varepsilon=3$. The array is illuminated by a plane wave with wavelength $\lambda=2.1{\mu}m$. In the forward FEM simulation we used a minimum element size of 0.6 nm with Perfectly Matching Layer (PML) boundary with $3\mu$m thickness, resulting in 2742129 total degrees of freedom. We used the electric field data shown in Fig. \ref{Fig2} (b) as the synthetic train dataset for PINNs.
In all simulations we sampled the electric field using a $150\times{150}$ spatial resolution. The utilized PINN architecture is composed of a feed-forward neural networks with 4 hidden layers and 64 neurons in each hidden layers. We used the hyperbolic tangent activation function and chose the Glorot uniform method for the weight initialization. The learning rate for the training was set to $10^{-3}$. Finally, we train the neural networks using the Adam methods and train for 150000 iterations (epochs) before reaching a satisfactory loss value of $10^{-2}$. The predicted $\varepsilon(x,y)$ profile obtained by this PINN is shown in Fig. \ref{Fig2} (c). We note that for the considered parameters of the scattering array PINN retrieves the permittivity distribution of a single dielectric nanocylinder with an effective dielectric function. Therefore, training PINNs to predict the permittivity distribution of the finite-size cluster of scattering cylinders that yield the same field distribution of the forward FEM scattering calculation has resulted in the effective homogenization of a finite-size metamaterial. 
Since we are working here under the conditions of small size parameter for the nanocylinders ($r,~a<<\lambda$) and weak scattering ($\varepsilon=3$), we can also estimate the effective medium permittivity using the classical Bruggeman mixing formula \cite{sihvola1999electromagnetic}:
\begin{equation}
\sum_{i} f_{i} \frac{\varepsilon_{i}-\varepsilon_{e}}{\varepsilon_{i}+ \varepsilon_{e}}=0,
\label{Bruggeman}
\end{equation}
where $\varepsilon_{e}$ is the effective permittivity, $f_{i}$ and $\varepsilon_{i}$ are the filling fraction and permittivity of each component, respectively. By using Eq. \ref{Bruggeman}, we predicted $\varepsilon_{e}=1.25$. The epsilon distribution $\varepsilon(x,y)$ from PINNs averaged inside the cylindrical windowed region ($r<5{\mu}m$) is $1.35\pm0.056$. The permittivity error between the PINN prediction and the Bruggeman model is $7\%$. Therefore, the value computed by the Bruggeman mixing formula in the weak scattering regime is compatible with the one predicted using PINNs. A direct validation of the PINN results is obtained by directly employing the $\varepsilon$ profile estimated by PINN and shown in Fig. \ref{Fig2} (c) inside forward scattering FEM simulations. The FEM computed total field $E_{z}$ distribution considering the PINN-predicted $\varepsilon$ under the same excitation conditions is shown in Fig. \ref{Fig2} (d). We further calculated the $L^2$ error norm between Fig. \ref{Fig2} (b) and (d). This error is $2.82\%$, indicating a very good retrieval of the permittivity profile using PINNs.
\begin{figure}[h!]
\centering\includegraphics[width=\textwidth]{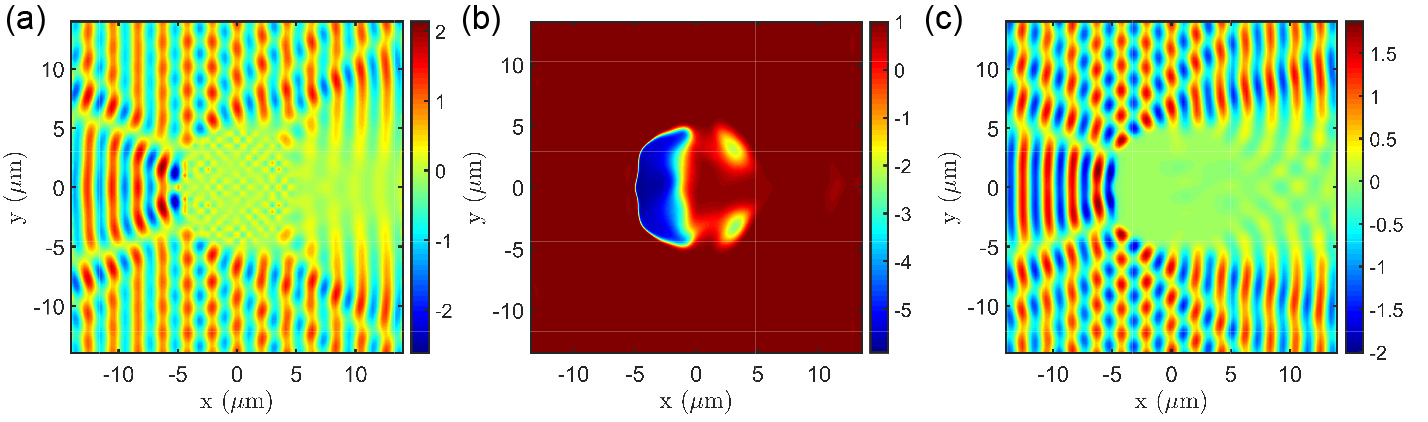}
\caption{(a) depicts the training dataset for the PINNs from FEM simulation of electric field real part distribution profile for square periodically arranged scatters with $a=500nm$, $r=125nm$, $d=10{\mu}m$, $N=317$, $\varepsilon=12$. The illumination plane wave wavelength is $\lambda=2.1{\mu}m$. (b) shows the PINNs predicted inverse permittivity profile by training with dataset shown in panel (a). (c) illustrates the FEM validation of electric field distribution when illuminating plane wave with wavelength $\lambda$ on the object with PINNs predicted permittivity profile.}\label{Fig3}
\end{figure}

To further study the role on the homogenization of the radiation effects introduced by the cylindrical scatterers we increased the permittivity of each nanocylinder in the array by considering $\varepsilon=12$. Fig. \ref{Fig3} (a) illustrates the real part of the total electric field distribution obtained from the plane wave excitation of such an array using FEM simulations with the same parameters as for the case in Fig. \ref{Fig2}. However, in the present case we observe a strong perturbation of the electric field around the array indicating the presence of stronger scattering effects. Using the same parameters and training method as in the previous case, PINN predicts now the permittivity profile displayed in Fig. \ref{Fig3} (b). Differently from the results previously shown in Fig. \ref{Fig2} (c), now PINN retrieves a non-homogeneous effective medium that contains a spatial region of negative effective permittivity. We implemented this permittivity profile in the FEM simulation and obtained the electric field distribution shown in Fig. \ref{Fig3} (c). The quality of the PINN prediction is again quantified by considering
the $L^2$ error norm between data in Fig. \ref{Fig3} (a) and (c), which is now $5\%$. Our findings indicate that PINNs effectively account for scattering and radiation effects in the array by retrieving a non-homogeneous effective medium with resonant permittivity profile, beyond the capabilities of traditional effective medium theory. 
\begin{figure}[h!]
\centering\includegraphics[width=10cm]{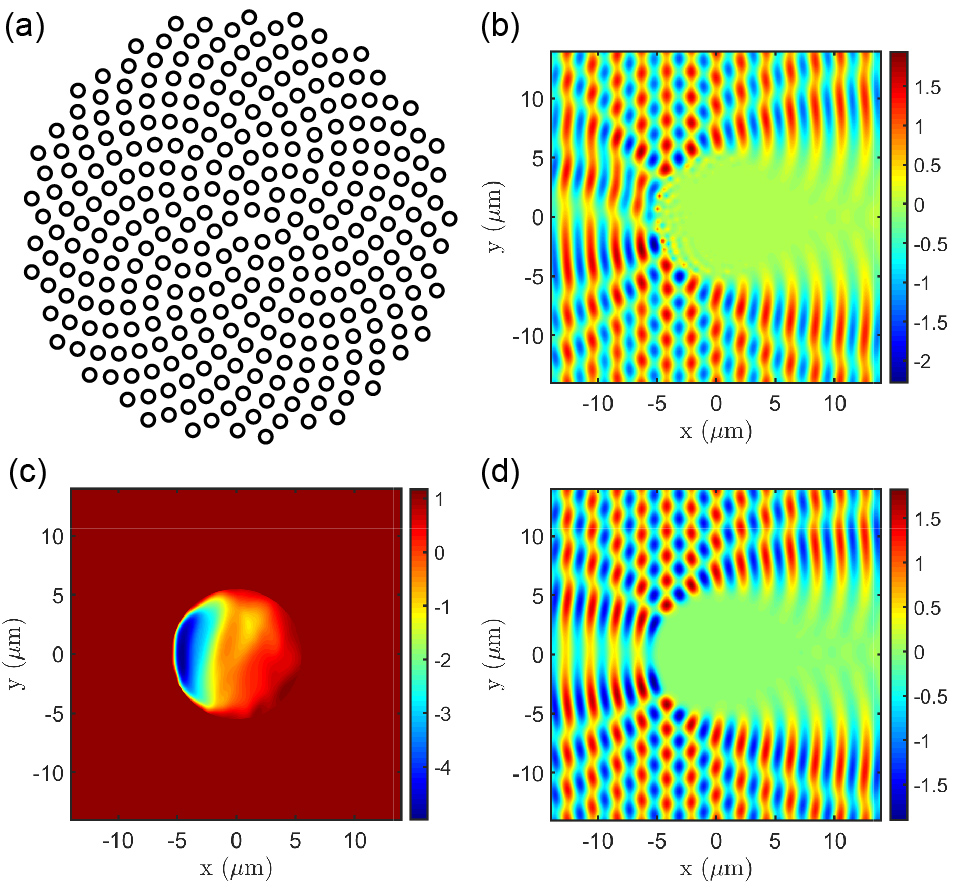}
\caption{FEM simulation of electric field real part distribution profile for Vogel spiral arranged scatters. The illumination plane wave wavelength is $\lambda=2.1{\mu}m$. (b) depicts the inverse train dataset for training PINNs homogenization from panel (a). (c) shows PINNs predicted inverse permittivity profile by training with dataset shown in panel (b). (d) illustrates the FEM validation of electric field distribution when illuminating plane wave with wavelength $\lambda$ on the object with PINNs predicted permittivity profile. The $L_2$ error between the panel (b) and (d) is $3.8\%$.}\label{Fig4}
\end{figure}

The proposed PINNs framework for metamaterial inverse design is not limited by the periodic arrangement of the scattering nanocylinders in the array. In fact, the exact same procedure can be utilized for the synthesis of non-homogeneous effective media corresponding to scattering arrays with arbitrary aperiodic morphology. In order to prove this point we consider next the case of a Vogel spiral array. Vogel spirals have been intensively investigated in plasmonics and nanophotonics literature thanks to their unique light scattering and localization properties that stimulated novel device applications \cite{trevino2012geometrical,lawrence2012control,pollard2009low,trevino2011circularly,liew2011localized,trevino2012plasmonic,razi2019optimization,fab2019localization}. 
Vogel spirals are defined in polar coordinates by following parametric equations:
\begin{equation}
\begin{aligned} 
\begin{cases}
r_{n} &=a_{0} \sqrt{n} \\ \theta_{n} &=n \alpha 
\end{cases}
\end{aligned}
\end{equation}\label{Vogel spiral}
where n = 0, 1, 2, ... is an integer, $a_0$ is a positive constant called scaling factor, and $\alpha$ is an irrational number, known as the divergence angle\cite{adam2011mathematical}. This angle specifies the constant aperture between successive point particles in the array. Since the divergence angle is an irrational number, Vogel spiral point patterns lack both translational and rotational symmetry.

Here we studied PINNs homogenization of a Vogel spiral array in the radiative (scattering) limit. The Vogel spiral pattern considered here is shown in Fig. \ref{Fig4} (a) and is characterized by an averaged center-to-center interparticle distance $\bar{a}=500nm$, $d=10{\mu}m$, $r=125nm$, $\varepsilon=12$, and $N=300$. The corresponding total electric field distribution computed with FEM under plane wave excitation with $\lambda=2.1{\mu}m$ is shown in Fig. \ref{Fig4} (b). We can appreciate immediately the asymmetric distribution of the total field around the Vogel spiral array and also the strong perturbation of the field due to the strong scattering effects of the nanocylinder. By training with the same network parameters established in previous case, PINN now predicts the effective permittivity distribution  $\varepsilon(x,y)$ illustrated in Fig. \ref{Fig4} (c). We note that the $\varepsilon$ profile recovered from PINN is asymmetric, which results from the asymmetric nature of the Vogel spiral structure. It also contains a negative permittivity region that effectively accounts for the strong radiation effects in the Vogel array. Finally, we show in Fig. \ref{Fig4} (d) the electric field distribution obtained from FEM forward simulations by using as an input the effective permittivity profile obtained from PINN. The $L^2$ error norm between the total electric fields shown in Fig. \ref{Fig4} (b) and (d) is found to be $3.8\%$. 

The proposed method can naturally be generalized to account for radiation losses in the multiple scattering medium by retrieving the complex permittivity of the effective medium.
In this case we need to consider Eq. \ref{inver homogenization} with both $E_z$ and $\epsilon_r(x,y)$ as complex variables. Separating real and imaginary parts, we then obtain the following PDE model for PINNs:
\begin{equation}
\begin{aligned} 
\begin{cases}
    \nabla^{2}Re\{E_z\}\left(x,y\right)+\left[Re\{E_z\}Re\{\varepsilon_{r}\left(x,y\right)\}-Im\{E_z\}Im\{\varepsilon_{r}\left(x,y\right)\}\right]k_{0}^{2}=0\\
    \nabla^{2}Im\{E_z\}\left(x,y\right)+\left[Im\{E_z\}Re\{\varepsilon_{r}\left(x,y\right)\}+Re\{E_z\}Im\{\varepsilon_{r}\left(x,y\right)\}\right]k_{0}^{2}=0
\end{cases}
\end{aligned}
\end{equation}\label{effective medium complex}
where $Re\{\cdot\}$ and $Im\{\cdot\}$ denote the real and imaginary parts of the corresponding quantities. We can now train PINNs using both the real and the imaginary parts of $E_{z}$ computed from the FEM simulations. This will enable PINNs to predict independently  $Re\{\varepsilon_{r}\left(x,y\right)\}$ and $Im\{\varepsilon_{r}\left(x,y\right)\}$ for the effective medium and therefore to quantify its radiation losses. We have implemented this more general model using the same neural network parameters and training method described in the previous cases and we have applied them to the structure shown in Fig. \ref{Fig2}$-$\ref{Fig4} in order to predict $Im\{\varepsilon_{r}\left(x,y\right)\}$ for the respective effective media. The maximum values of $Im\{\varepsilon_{r}\left(x,y\right)\}$ obtained in these three cases are $10^{-4}$, $0.6$, and $0.3$, respectively. Finally we notice that the real parts $Re\{\varepsilon_{r}\left(x,y\right)\}$ of the predicted complex permittivity remain very close, within a total $3\%$ error, to the values previously obtained considering only the real part of the effective medium. However, the more general approach outlined above allows one to investigate arrays of nanocylinders with even larger refractive index contrast where radiation losses, which are difficult to account using traditional effective index theory \cite{sihvola1999electromagnetic}, are expected to play a very important role.

Therefore, our results demonstrate that the general PINNs framework is suitable to systematically study the effects of morphology and radiation coupling in arbitrary non-periodic effective media and metamaterials, beyond the limitations of effective medium theory. The ability to reliably identify the effective media that yield the same total field as arbitrary scattering arrays using PINNs can not only greatly stimulate the development of novel finite-size metamaterials and resonant nanostructures but also provides fundamental insights to advance the current status of advanced homogenization theory with radiation effects \cite{rechtsman_effective_2008,kim_multifunctional_2019}.

\section{PINN for inverse Mie scattering}
We showed in previous section that PINNs are capable of homogenizing arrays composed of sub-wavelength particles arranged in arbitrary aperiodic geometries. In this section we will further show that, based on the knowledge of the field values around compound optical nanostructures of known morphology, PINNs can successfully retrieve their optical parameters. This specific application requires the solution of an inverse problem with a PINN that, differently from the previous sections, includes the specific boundary conditions in the definition of the appropriate PDE model. The results shown in this section demonstrate that PINNs framework can potentially be applied to retrieve the optical properties of complex nanostructures from near-field imaging data \cite{novotny2012principles}. 

\begin{figure}[h!]
\centering\includegraphics[width=10cm]{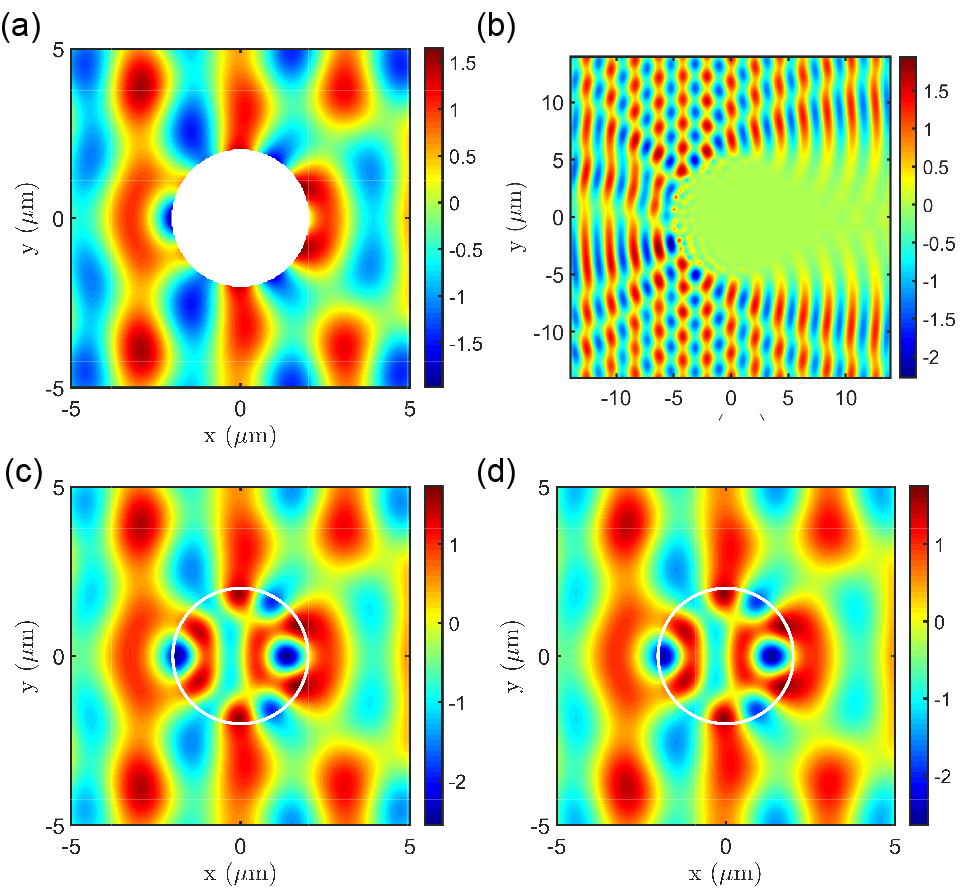}
\caption{(a) shows the train dataset for inverse Mie scattering theory by using PINNs; we used only the external electric field distribution with the nanocylinder radius $r=2{\mu}m$, $\varepsilon_{r2}=4$, incident wavelength $\lambda=3{\mu}m$, (b) shows PINNs inverse finding of the nanocylinder permittivity as a function of iteration step, (c) shows the reconstructed electric field distribution for the nanocylinder scattering from PINNs, (d) illustrates the real electric field distribution of the given nanocylinder using Mie theory. The $L_2$ error between the panel (c) and (d) is $0.51\%$.}\label{Fig5}
\end{figure}

We start our analysis by considering the case of retrieving the permittivity of a single resonant nanocylinder from its external total field. We compute the external electric field distribution, shown in Fig. \ref{Fig5} (a), using the analytical Mie scattering theory \cite{bohren2008absorption} for a cylinder with radius $r=2{\mu}m$ and permittivity $\varepsilon=4$ under TM wave excitation at a wavelength $\lambda=3{\mu}m$. Note that the field inside the cylinder is not specified here. In order to retrieve the internal field distribution along with the unknown permittivity parameter of the nanocylinder, we implemented the following PDE model:
\begin{equation}\label{inverse single Mie}
\begin{aligned}
\begin{cases}
\nabla^{2} E_{z}^{(k)}+\varepsilon_{rk}k_{0}^{2} E_{z}^{(k)}=0 \quad in~\Omega_{k},(k=1,2)\\
\left.E_{z}^{(1)}\right|_{r=a}=\left.E_{z}^{(2)}\right|_{r=a}\\
\left.\frac{1}{\mu_{r1}}\frac{\partial E_{z}^{(1)}}{\partial r}\right|_{r=a}=\left.\frac{1}{\mu_{r2}}\frac{\partial E_{z}^{(2)}}{\partial r}\right|_{r=a}
\end{cases}
\end{aligned}
\end{equation}
where $E_{z}^{(1)}$, $E_{z}^{(2)}$ denotes the electric field $z$ component real part inside and outside the nanocylinder, respectively. Here $r$ denotes the radial components in cylindrical coordinates while $a$ is the nanocylinder radius. $\varepsilon_{r1}$, $\varepsilon_{r2}$ are the relative permittivity of the nanocylinder and the host medium, respectively. $\mu_{r1}$, $\mu_{r2}$ are the relative permeability of the corresponding regions. We are considering non-magnetic materials in this section so we have $\mu_{r1}=\mu_{r2}=1$. Moreover, we set $\varepsilon_{r2}=1$ while $\varepsilon_{r1}$ is a trainable variable that PINNs retrieve through the training process fed by the analytically computed external field.

The same network architecture and hyperparameters as in the previous training cases are utilized here. However, we multiplied the loss term $\mathcal{T}_{i}$ by a factor 100 in order to enforce a stronger penalty and obtain better train results. We stop training after $10^{5}$ steps that we found sufficient to achieve a $10^{-4}$ train loss value. The retrieval of $\varepsilon_{r2}$ during the training process is illustrated in Fig. \ref{Fig5} (b), where the dashed line indicates the true solution from analytical Mie theory. We can see clearly that PINNs successfully retrieve the true solution when fed with only information on the external field. Furthermore, the reconstructed field from PINNs, displayed in Fig. \ref{Fig5} (c), shows a very small difference from the Mie theory (Fig. \ref{Fig5} (d)) that is quantified by an $L^2$ error norm of $0.51\%$. Therefore, we show that PINNs can successfully retrieve both the internal field and the cylinder permittivity when constrained by a physical PDE model. 
\begin{figure}[h!]
\centering\includegraphics[width=10cm]{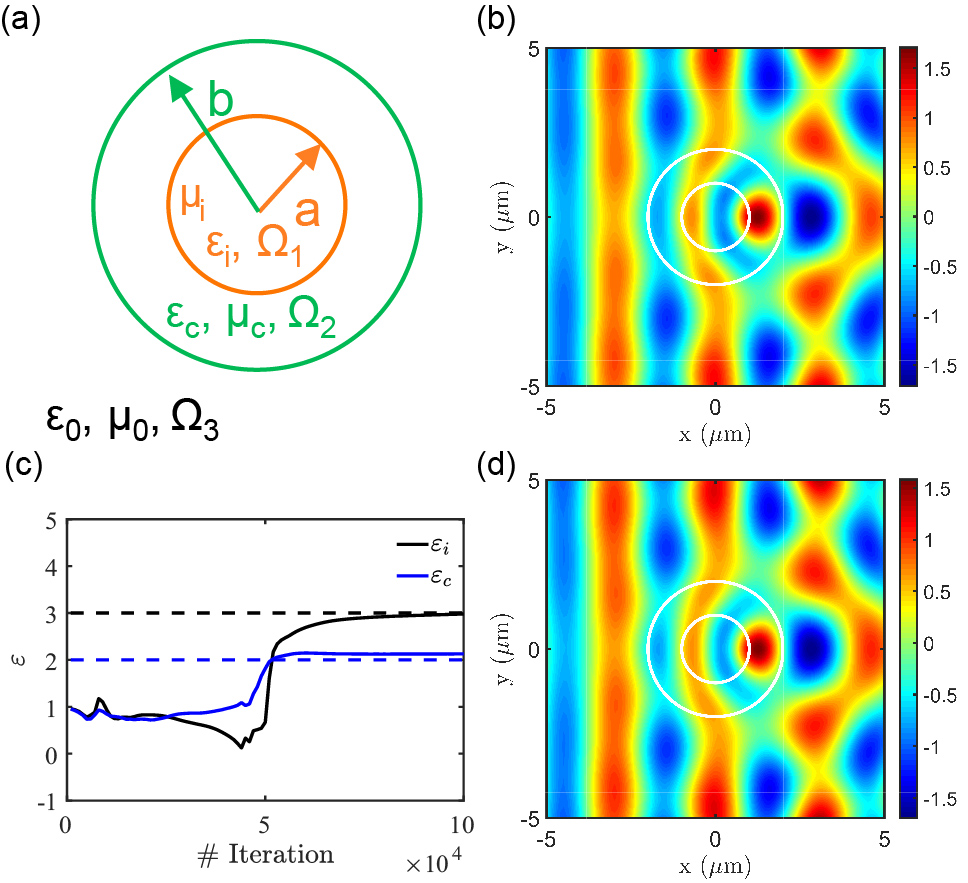}
\caption{(a) shows the schematic of the studied coated nanocylinder parameters, (b) shows reconstructed electric field distribution for the coated nanocylinder under the excitation of plane wave with $\lambda=3{\mu}m$, (c) illustrates the inverse finding of the two parameters $\epsilon_{i}$, $\epsilon_{c}$ with respect to the number of iteration of PINN, (d) shows the reconstructed electric field distribution from PINNs after $10^{4}$ training steps.}\label{Fig6}
\end{figure}

Furthermore, we now extend PINNs to the retrieval of multiple materials parameters by considering the case of a coated nanocylinder. The geometry of the studied coated nanocylinder is schematically illustrated in Fig. \ref{Fig6} (a). The spatial distribution of the real part of the total electric field of the coated nanocylinder with $a=0.5{\mu}m$, $b=1{\mu}m$, $\varepsilon_{i}=3$, $\varepsilon_{c}=2$ is shown in Fig. \ref{Fig6} (b). The corresponding PDE model that we implemented within PINNs framework to solve this inverse problem is reported below: 
\begin{equation}\label{inverse Mie coated}
\begin{aligned}
\begin{cases}
\nabla^{2} E_{z}^{(k)}+\varepsilon_{k}k_{0}^{2} E_{z}^{(k)}=0 \quad in~\Omega_{k},(k=1,2,3)\\
\left.E_{z}^{(1)}\right|_{r=a}=\left.E_{z}^{(2)}\right|_{r=a},\quad\left.E_{z}^{(2)}\right|_{r=b}=\left.E_{z}^{(3)}\right|_{r=b}\\
\left.\frac{1}{\mu_{i}}\frac{\partial E_{z}^{(1)}}{\partial r}\right|_{r=a}=\left.\frac{1}{\mu_{c}}\frac{\partial E_{z}^{(2)}}{\partial r}\right|_{r=a},\quad\left.\frac{1}{\mu_{c}}\frac{\partial E_{z}^{(2)}}{\partial r}\right|_{r=b}=\left.\frac{1}{\mu_{0}}\frac{\partial E_{z}^{(3)}}{\partial r}\right|_{r=b}
\end{cases}
\end{aligned}
\end{equation}
where $E_{z}^{(1)}$, $E_{z}^{(2)}$, $E_{z}^{(3)}$ denote the $z$ components of the real parts of the electric fields inside the inner core, coated layer, and outside coating layer, respectively. $\varepsilon_{i}$, $\varepsilon_{c}$, $\varepsilon_{0}$ are the relative permittivities of the inner-core, coated layer, and host medium, respectively. $\mu_{i}$, $\mu_{c}$, $\mu_{0}$ are the relative magnetic permeabilities of the corresponding regions that are here set equal to unity since we are considering non-magnetic materials. In the parameter retrieval process, we fix $\varepsilon_{0}=1$ and train PINNs to predict $\varepsilon_{i}$ and $\varepsilon_{c}$ based on the field distribution from Fig. \ref{Fig6} (b).  Specifically, We utilize the same neural network architecture and training method as for the single nanocylinder case. Figure \ref{Fig6} (c) shows clearly the parameter retrieval during the training process, which converges to the exact values used to create the training data. Figure \ref{Fig6} (d) shows the reconstructed field. The $L^2$ error norm between Fig. \ref{Fig6} (b) and (d) is only $1\%$. Therefore, we showed that PINNs can successfully retrieve multiple material parameters for compound resonant nanostructures uniquely based on the total field information surrounding the objects.
\begin{figure}[h!]
\centering\includegraphics[width=10cm]{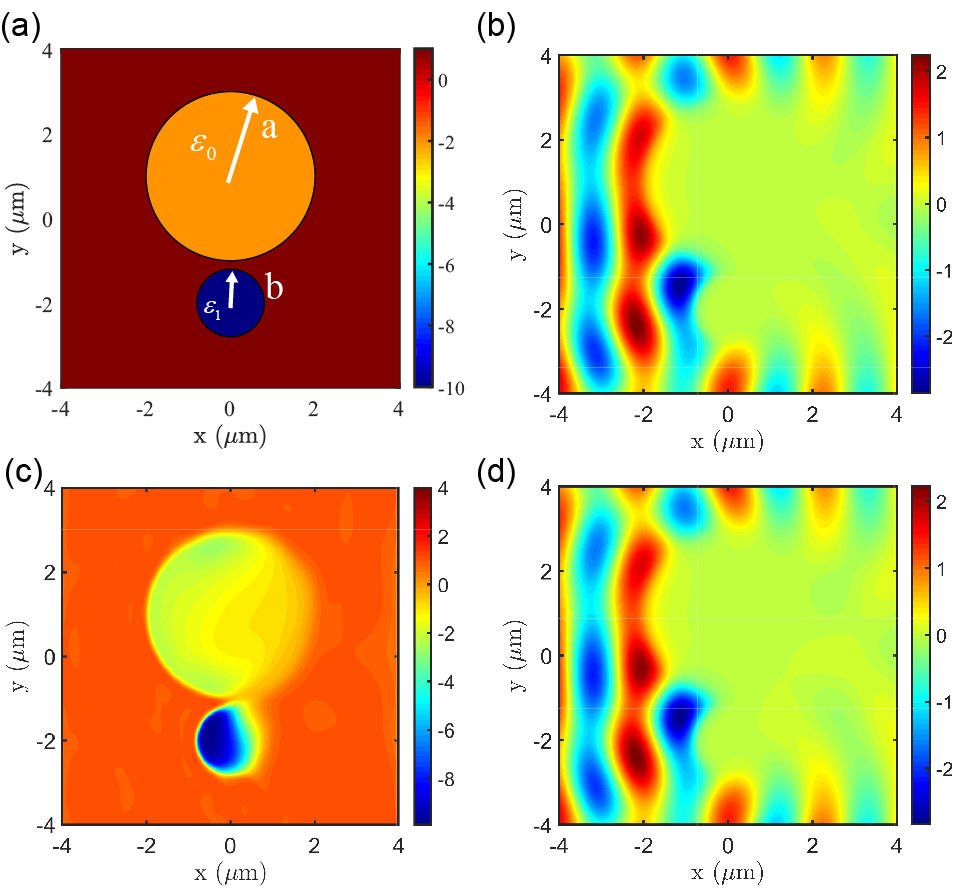}
\caption{(a) shows the permittivity profile of the two asymmetric dimer with $\varepsilon_0=-2$, $\varepsilon_1=-10$, $r_0=2{\mu}m$, $r_1=0.8{\mu}m$, (b) depicts train dataset for PINN from FEM simulation of electric field distribution profile for the asymmetric dimer shown in (a). The incident electromagnetic wave wavelength is $\lambda=2.1{\mu}m$ (c) shows the PINN predicted permittivity profile after train, (d) illustrates the FEM validation of the electric field distribution for predicted permittivity by PINN. The $L_2$ error between the panel (b) and (d) is $0.30\%$.}\label{Fig7}
\end{figure}

We show next that in addition to retrieving the optical parameters of individual nanostrucutres, PINNs can additionally extract the parameters in systems composed of multiple objects. Inspired from the scattering target FoamDielExtTM \cite{geffrin2005free} frequently used in diffraction tomography at GHz frequencies \cite{sun2018efficient,liu2017seagle,kamilov2017plug}, we study parameter retrieval in the setup shown in Fig. \ref{Fig7} (a). The real part of the total electric field distribution for the considered structure under incident plane wave illumination at wavelength $\lambda=2.1{\mu}m$ is illustrated in Fig. \ref{Fig7} (b). Using this field as the train dataset, we couple PINNs with the PDE homogenization model in Eq.\ref{inver homogenization} and directly retrieve the permittivity profile over the entire computational window that gives rise to the same total field as the asymmetric dimer in \ref{Fig7} (a). In order to improve the quality of the reconstruction here we utilize the ResNet architecture with two residual blocks and each residual block has two fully-connected layers with width 64. We used the same activation function and initialization strategy as in all previous simulations. We multiply again the loss term $\mathcal{T}_{i}$ by a factor 100 to obtain better train results. The training process uses the Adam optimizer and is stopped after 150000 iterations resulting in $10^{-4}$ training losses. The retrieved $\varepsilon(x,y)$ profile from this PINN is shown in Fig. \ref{Fig7} (c). We observe that the PINN reconstructed two object permittivities localized within regions that qualitatively correspond to the input target. The electric field distribution obtained by FEM using the permittivity profile retrieved by PINN is illustrated in Fig. \ref{Fig7} (d), where the $L^2$ error norm between these two field profiles is $0.3\%$. Notice that in the PDE homogenization model used here there are no boundary conditions specified to constrain the PINN solution. As a result, PINN found an equivalent permittivity profile that gives rise to the same field distribution as the train dataset but this solution may not be unique. This is the reason why the retrieved permittivity distribution is quite different in this case from the input target, despite the two share the same scattering behavior within the considered window.
In all cases, when the morphology of a scattering object is known or characterized experimentally, PINNs framework can be utilized to retrieve nanomaterial properties based on scattered fields that can, for instance, be measured from scanning near-field optical microscopy (SNOM) \cite{hecht2000scanning}. The ability to infer material properties directly from imaging data using PINNs can provide unique opportunities for remote sensing and detection in the optical regime. 

\section{PINN for invisible cloaking design}
\begin{figure}[h!]
\centering\includegraphics[width=10cm]{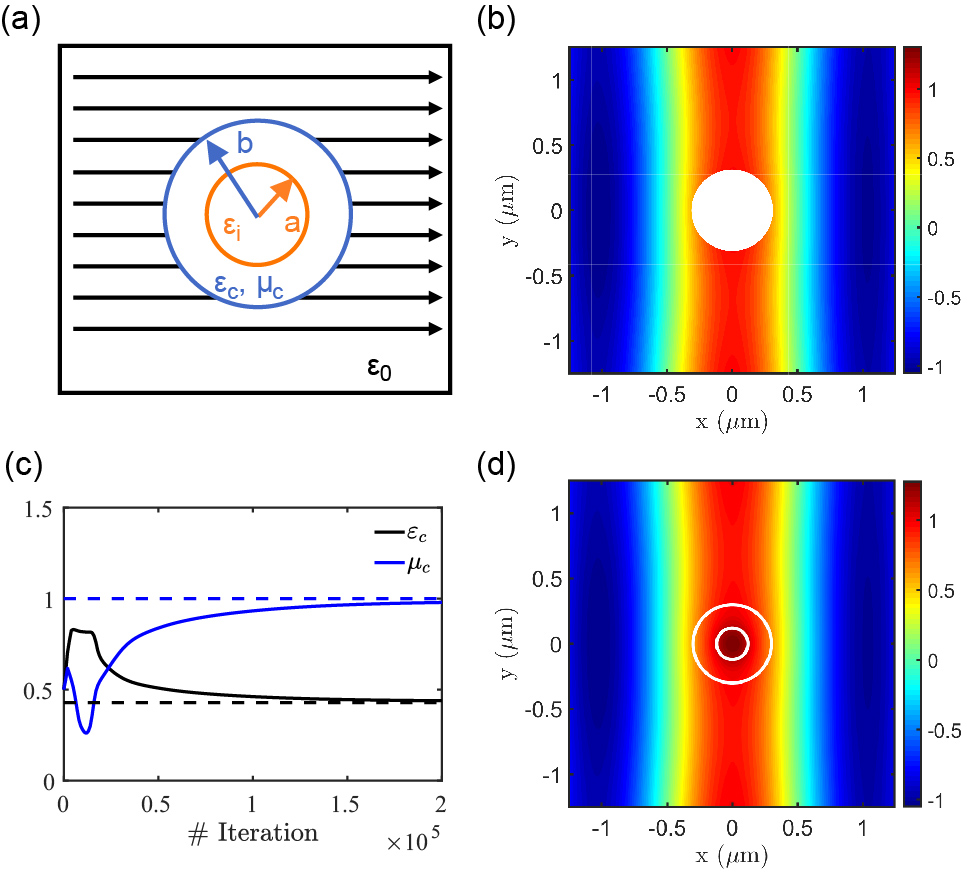}
\caption{(a) shows the schematic for nanocylinder with constant permittivity coated layer zeroing out scattering, $a=0.12{\mu}m$, $b=0.3{\mu}m$, $\epsilon_i=4$, (b) shows the inverse train dataset for PINNs, the excitation plane wave wavelength $\lambda=2.1{\mu}m$, (c) illustrates PINNs predicted permittivity for the coating layer to zero out scattering. The dashed line is the analytical solution of the coating layer permittivity to zero out the nanocylinder scattering. (d) demonstrates the electric field distribution for coated nanocylinder with the coating layer permittivity predicted by PINN.}\label{Fig8}
\end{figure}

In this section, we will show how to reformulate the well-known invisible cloaking problem into a general parameter retrieval problem that can be accurately solved using PINNs. The cloaking problem has stimulated significant interest in recent years \cite{alu2008plasmonic,fleury2015invisibility}, and admits an analytical solution in the limit of small objects compared to the incoming wavelength (quasi-static limit). The considered geometry is illustrated in Fig. \ref{Fig8} (a) and consists of an inner nanocylinder with radius $a$ and permittivity $\varepsilon_{i}$ coated by a cloaking material with thickness equals to $b-a$ and unknown permittivity $\varepsilon_{c}$ that cancels the scattering for a given plane wave incidence condition. We chose the following parameters for our cloaking example: $a=0.12{\mu}m$, $b=0.3{\mu}m$, $\epsilon_i=4$.  Notice that for nanocylinders of small size parameters ($k_0b<<1$), we can obtain the permittivity of the cloaking layer $\varepsilon_{c}$ using the following analytical formula \cite{alu2010plasmonic}:
\begin{equation}\label{analytical permittivity for invisible cloaking}
\frac{b}{a}=\sqrt{\frac{\varepsilon_{\mathrm{c}}-\varepsilon_{i}}{\varepsilon_{\mathrm{c}}-\varepsilon_{0}}}
\end{equation}
enabling verification of PINN-predicted results. 

In order to generate the training data for PINNs we computed the electric field distribution of the coated nanocylinder using Mie theory with $\varepsilon_{\mathrm{c}}$ that satisfies the cloaking Eq. \ref{analytical permittivity for invisible cloaking}. The obtained field distribution (excluding the internal field) is then used as the train dataset for PINNs as illustrated in Fig. \ref{Fig8} (b). In this example we used the PDE model previously defined by Eq. \ref{inverse Mie coated}. We again employ a simple feed-forward neural network with 4 hidden layers and 64 neurons in each hidden layers. Hyperbolic tangent activation function and Glorot uniform initializer are used as in previous cases. However, here we utilized a learning rate for the training equal to $10^{-4}$. We trained the PINN neural networks with the Adam optimizer and considered $2\times{10^{5}}$ iteration steps, resulting in overall training losses equal to $10^{-4}$. In order to obtain better accuracy we multiply the loss term $\mathcal{T}_{i}$ as well as the loss term from the $E_z$ continuity at interface $r=b$ by a factor 100. We let the PINN retrieves both the electric permittivity and the magnetic permeability. The parameter retrieval curves as a function of the iteration number are shown in Fig. \ref{Fig8} (c). These data demonstrate that PINNs correctly find the true solution of the coating layer permittivity in complete agreement with the analytical expression. Moreover, it is important to note that the considered PINN obtains a magnetic (relative) permittivity equal to unity, confirming that no magnetic response is necessary for perfect scattering cancellation in this geometry for a small particle size. Fig. \ref{Fig8} (d) displays the reconstructed electric field from PINN, showing clearly that the electric field propagates the coated nanocylinder without any perturbation. 

\begin{figure}[h!]
\centering\includegraphics[width=12cm]{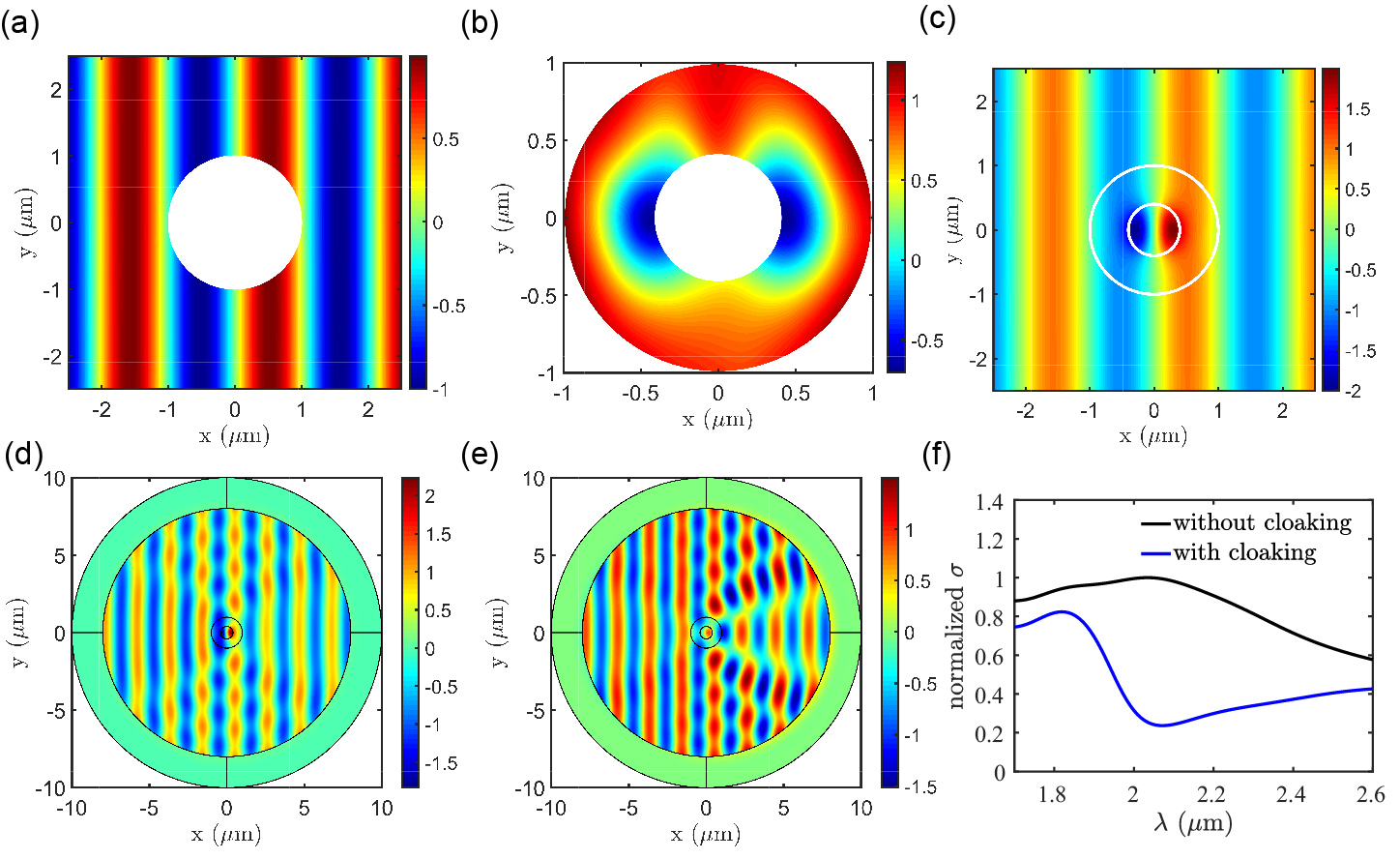}
\caption{(a) shows the inverse train dataset for PINNs, where the excitation plane wave wavelength $\lambda=2.1{\mu}m$, (b) illustrates PINNs predicted permittivity distribution for the coating layer, (c) depicts the field distribution reconstructed from PINNs for the coated nanocylinder, (d), (e) show the electric field distribution for coated nanocylinder and without coating nanocylinder, respectively. (f) illustrates the normalized scattering cross section $\sigma$ spectrum for the nanocylinder with or without PINNs identified cloaking.}\label{Fig9}
\end{figure}

Finally, we address the conditions to achieve radiation cloaking in coated nanocylinders whose diameter equals the incoming wavelength. Although no perfect cloaking condition is theoretically expected in this case, it is still very interesting to use PINNs to discover the coating parameters for the optimal scattering reduction. The geometry of the problem is the same as in Fig. \ref{Fig8} (a) but here we additionally let the permittivity of the coating layer to be spatially dependent $\varepsilon_{c} (x,y)$ (within the layer) in order to obtain better cloaking results. Figure \ref{Fig9} (a) shows the training dataset for PINN under plane wave illumination with wavelength $\lambda=2.1{\mu}m$, and corresponds to a wave that propagates through the coated nanocylinder undisturbed. We emphasize that this situation cannot physically occur when the particle size is comparable to the wavelength. However, we will show that training under such a desirable condition allows PINNs to discover a permittivity profile that substantially reduces the scattering cross section of the cylinder. 

We used the same PDE model as defined by Eq. \ref{inverse Mie coated} but we fixed the value to the magnetic permeability to unity in order to design a purely dielectric device. Given the complexity of this problem we modified the neural network architecture by considering 5 hidden layers while keeping all the other parameters as in previous cases. We additionally multiply the loss term $\mathcal{T}_{i}$ by a factor of 20 in order to obtain better training quality. The neural network is trained up to $2\times{10^{5}}$ steps and training loss values around $10^{-4}$ are achieved in the end. 

Figures \ref{Fig9} (b) and (c) display the PINN-predicted $\epsilon_{c}(x,y)$ distribution and the total field distribution. The PINN-predicted $\epsilon_{c}(x,y)$ is then used in the FEM simulations resulting in the total electric field profile shown in Fig. \ref{Fig9} (d). For comparison, we also show the electric field distribution for the bare nanocylinder without the coating layer in Fig. \ref{Fig9} (e). It is clear that PINNs identified the dielectric properties of the coating layers that significantly reduce the far-field scattering of the device under plane wave excitation. Specifically, we quantified this by computing the far-field scattering efficiency of the coated cylinder and confirmed a $75\%$ reduction in the scattering efficiency due to the coating layer characteristics identified using PINNs, as shown in Fig. \ref{Fig9} (f). In summary, using the flexible PINNs framework we have established not only a powerful approach to retrieve material parameters based on scattering data (synthetic or measured) but also to design novel dielectric devices with strongly reduced scattering properties. 

\section{Conclusions}
{We demonstrated in this paper the solution of representative inverse scattering problems that are particularly interesting in photonic metamaterials and nano-optics technologies. In particular, we successfully introduced and validated the PINNs framework for the effective medium reconstruction of scattering arrays where finite-size and radiation effect significantly perturb the classical homogenization picture. Our approach leverages recent advances in deep-learning algorithms constrained by the wave physics of the problems, which dramatically simplifies data training procedures compared to alternative machine learning approaches. Our findings have been fully validated using Finite Element Method (FEM) numerical simulations. In addition, we developed the PINNs framework for the retrieval of multiple optical parameters based on the field distribution around nanomaterials, which suggests a method to directly access material information from near-field imaging data. Finally, we use PINNs to address the problem of invisible cloaking with coated cylinders and show $75\%$ scattering abatement in dielectric coated cylinders with size comparable to the wavelength of the incoming radiation. 
We believe that further PINNs development for inverse scattering and remote sensing of nanostructures can significantly broaden current design capabilities and also provide fundamental insights to advance the current state of homogenization theory with radiation effects. Although the PINNs architectures employed here are based on empirical tuning, recent advances in meta-learning \cite{vanschoren2018metalearning,chen2019learning,he2019automl} can enable future automated selection of the optimum architectures.} 

\section*{Funding}
{This research was sponsored by the Army Research Laboratory and was accomplished under Cooperative Agreement Number W911NF-12-2-0023. The views and conclusions contained in this document are those of the authors and should not be interpreted as representing the official policies, either expressed or implied, of the Army Research Laboratory or the U.S. Government. The U.S. Government is authorized to reproduce and distribute reprints for Government purposes notwithstanding any copyright notation herein. This work was also supported by the DOE PhILMs project (No. de-sc0019453).}

\section*{Disclosures}
The authors declare no conflict of interest.

\section*{Acknowledgments}
L. D. N. would like to thank Professors Nader Engheta and Salvatore Torquato for insightful discussions on non-homogeneous metamaterials and effective medium theory.

%


\begin{thebibliography}{40}%
\makeatletter
\providecommand \@ifxundefined [1]{%
 \@ifx{#1\undefined}
}%
\providecommand \@ifnum [1]{%
 \ifnum #1\expandafter \@firstoftwo
 \else \expandafter \@secondoftwo
 \fi
}%
\providecommand \@ifx [1]{%
 \ifx #1\expandafter \@firstoftwo
 \else \expandafter \@secondoftwo
 \fi
}%
\providecommand \natexlab [1]{#1}%
\providecommand \enquote  [1]{``#1''}%
\providecommand \bibnamefont  [1]{#1}%
\providecommand \bibfnamefont [1]{#1}%
\providecommand \citenamefont [1]{#1}%
\providecommand \href@noop [0]{\@secondoftwo}%
\providecommand \href [0]{\begingroup \@sanitize@url \@href}%
\providecommand \@href[1]{\@@startlink{#1}\@@href}%
\providecommand \@@href[1]{\endgroup#1\@@endlink}%
\providecommand \@sanitize@url [0]{\catcode `\\12\catcode `\$12\catcode
  `\&12\catcode `\#12\catcode `\^12\catcode `\_12\catcode `\%12\relax}%
\providecommand \@@startlink[1]{}%
\providecommand \@@endlink[0]{}%
\providecommand \url  [0]{\begingroup\@sanitize@url \@url }%
\providecommand \@url [1]{\endgroup\@href {#1}{\urlprefix }}%
\providecommand \urlprefix  [0]{URL }%
\providecommand \Eprint [0]{\href }%
\providecommand \doibase [0]{http://dx.doi.org/}%
\providecommand \selectlanguage [0]{\@gobble}%
\providecommand \bibinfo  [0]{\@secondoftwo}%
\providecommand \bibfield  [0]{\@secondoftwo}%
\providecommand \translation [1]{[#1]}%
\providecommand \BibitemOpen [0]{}%
\providecommand \bibitemStop [0]{}%
\providecommand \bibitemNoStop [0]{.\EOS\space}%
\providecommand \EOS [0]{\spacefactor3000\relax}%
\providecommand \BibitemShut  [1]{\csname bibitem#1\endcsname}%
\let\auto@bib@innerbib\@empty
\bibitem [{\citenamefont {Kamilov}\ \emph {et~al.}(2015)\citenamefont
  {Kamilov}, \citenamefont {Papadopoulos}, \citenamefont {Shoreh},
  \citenamefont {Goy}, \citenamefont {Vonesch}, \citenamefont {Unser},\ and\
  \citenamefont {Psaltis}}]{kamilov2015learning}%
  \BibitemOpen
  \bibfield  {author} {\bibinfo {author} {\bibfnamefont {Ulugbek~S}\
  \bibnamefont {Kamilov}}, \bibinfo {author} {\bibfnamefont {Ioannis~N}\
  \bibnamefont {Papadopoulos}}, \bibinfo {author} {\bibfnamefont {Morteza~H}\
  \bibnamefont {Shoreh}}, \bibinfo {author} {\bibfnamefont {Alexandre}\
  \bibnamefont {Goy}}, \bibinfo {author} {\bibfnamefont {Cedric}\ \bibnamefont
  {Vonesch}}, \bibinfo {author} {\bibfnamefont {Michael}\ \bibnamefont
  {Unser}}, \ and\ \bibinfo {author} {\bibfnamefont {Demetri}\ \bibnamefont
  {Psaltis}},\ }\bibfield  {title} {\enquote {\bibinfo {title} {Learning
  approach to optical tomography},}\ }\href@noop {} {\bibfield  {journal}
  {\bibinfo  {journal} {Optica}\ }\textbf {\bibinfo {volume} {2}},\ \bibinfo
  {pages} {517--522} (\bibinfo {year} {2015})}\BibitemShut {NoStop}%
\bibitem [{\citenamefont {Molesky}\ \emph {et~al.}(2018)\citenamefont
  {Molesky}, \citenamefont {Lin}, \citenamefont {Piggott}, \citenamefont {Jin},
  \citenamefont {Vuckovi{\'c}},\ and\ \citenamefont
  {Rodriguez}}]{molesky2018inverse}%
  \BibitemOpen
  \bibfield  {author} {\bibinfo {author} {\bibfnamefont {Sean}\ \bibnamefont
  {Molesky}}, \bibinfo {author} {\bibfnamefont {Zin}\ \bibnamefont {Lin}},
  \bibinfo {author} {\bibfnamefont {Alexander~Y}\ \bibnamefont {Piggott}},
  \bibinfo {author} {\bibfnamefont {Weiliang}\ \bibnamefont {Jin}}, \bibinfo
  {author} {\bibfnamefont {Jelena}\ \bibnamefont {Vuckovi{\'c}}}, \ and\
  \bibinfo {author} {\bibfnamefont {Alejandro~W}\ \bibnamefont {Rodriguez}},\
  }\bibfield  {title} {\enquote {\bibinfo {title} {Inverse design in
  nanophotonics},}\ }\href@noop {} {\bibfield  {journal} {\bibinfo  {journal}
  {Nature Photonics}\ }\textbf {\bibinfo {volume} {12}},\ \bibinfo {pages}
  {659--670} (\bibinfo {year} {2018})}\BibitemShut {NoStop}%
\bibitem [{\citenamefont {Liu}\ \emph {et~al.}(2017)\citenamefont {Liu},
  \citenamefont {Liu}, \citenamefont {Mansour}, \citenamefont {Boufounos},
  \citenamefont {Waller},\ and\ \citenamefont {Kamilov}}]{liu2017seagle}%
  \BibitemOpen
  \bibfield  {author} {\bibinfo {author} {\bibfnamefont {Hsiouyuan}\
  \bibnamefont {Liu}}, \bibinfo {author} {\bibfnamefont {Dehong}\ \bibnamefont
  {Liu}}, \bibinfo {author} {\bibfnamefont {Hassan}\ \bibnamefont {Mansour}},
  \bibinfo {author} {\bibfnamefont {Petros~T}\ \bibnamefont {Boufounos}},
  \bibinfo {author} {\bibfnamefont {Laura}\ \bibnamefont {Waller}}, \ and\
  \bibinfo {author} {\bibfnamefont {Ulugbek~S}\ \bibnamefont {Kamilov}},\
  }\bibfield  {title} {\enquote {\bibinfo {title} {Seagle: Sparsity-driven
  image reconstruction under multiple scattering},}\ }\href@noop {} {\bibfield
  {journal} {\bibinfo  {journal} {IEEE Transactions on Computational Imaging}\
  }\textbf {\bibinfo {volume} {4}},\ \bibinfo {pages} {73--86} (\bibinfo {year}
  {2017})}\BibitemShut {NoStop}%
\bibitem [{\citenamefont {Kamilov}\ \emph {et~al.}(2017)\citenamefont
  {Kamilov}, \citenamefont {Mansour},\ and\ \citenamefont
  {Wohlberg}}]{kamilov2017plug}%
  \BibitemOpen
  \bibfield  {author} {\bibinfo {author} {\bibfnamefont {Ulugbek~S}\
  \bibnamefont {Kamilov}}, \bibinfo {author} {\bibfnamefont {Hassan}\
  \bibnamefont {Mansour}}, \ and\ \bibinfo {author} {\bibfnamefont {Brendt}\
  \bibnamefont {Wohlberg}},\ }\bibfield  {title} {\enquote {\bibinfo {title} {A
  plug-and-play priors approach for solving nonlinear imaging inverse
  problems},}\ }\href@noop {} {\bibfield  {journal} {\bibinfo  {journal} {IEEE
  Signal Processing Letters}\ }\textbf {\bibinfo {volume} {24}},\ \bibinfo
  {pages} {1872--1876} (\bibinfo {year} {2017})}\BibitemShut {NoStop}%
\bibitem [{\citenamefont {Pham}\ \emph {et~al.}(2018)\citenamefont {Pham},
  \citenamefont {Soubies}, \citenamefont {Goy}, \citenamefont {Lim},
  \citenamefont {Soulez}, \citenamefont {Psaltis},\ and\ \citenamefont
  {Unser}}]{Pham2018Versatile}%
  \BibitemOpen
  \bibfield  {author} {\bibinfo {author} {\bibfnamefont {Thanh-An}\
  \bibnamefont {Pham}}, \bibinfo {author} {\bibfnamefont {Emmanuel}\
  \bibnamefont {Soubies}}, \bibinfo {author} {\bibfnamefont {Alexandre}\
  \bibnamefont {Goy}}, \bibinfo {author} {\bibfnamefont {Joowon}\ \bibnamefont
  {Lim}}, \bibinfo {author} {\bibfnamefont {Ferr\'{e}ol}\ \bibnamefont
  {Soulez}}, \bibinfo {author} {\bibfnamefont {Demetri}\ \bibnamefont
  {Psaltis}}, \ and\ \bibinfo {author} {\bibfnamefont {Michael}\ \bibnamefont
  {Unser}},\ }\bibfield  {title} {\enquote {\bibinfo {title} {Versatile
  reconstruction framework for diffraction tomography with intensity
  measurements and multiple scattering},}\ }\href@noop {} {\bibfield  {journal}
  {\bibinfo  {journal} {Optics Express}\ }\textbf {\bibinfo {volume} {26}},\
  \bibinfo {pages} {2749--2763} (\bibinfo {year} {2018})}\BibitemShut {NoStop}%
\bibitem [{\citenamefont {Colton}\ and\ \citenamefont
  {Kress}(2018)}]{colton2018looking}%
  \BibitemOpen
  \bibfield  {author} {\bibinfo {author} {\bibfnamefont {David}\ \bibnamefont
  {Colton}}\ and\ \bibinfo {author} {\bibfnamefont {Rainer}\ \bibnamefont
  {Kress}},\ }\bibfield  {title} {\enquote {\bibinfo {title} {Looking back on
  inverse scattering theory},}\ }\href@noop {} {\bibfield  {journal} {\bibinfo
  {journal} {SIAM Review}\ }\textbf {\bibinfo {volume} {60}},\ \bibinfo {pages}
  {779--807} (\bibinfo {year} {2018})}\BibitemShut {NoStop}%
\bibitem [{\citenamefont {Sun}\ \emph {et~al.}(2018)\citenamefont {Sun},
  \citenamefont {Xia},\ and\ \citenamefont {Kamilov}}]{sun2018efficient}%
  \BibitemOpen
  \bibfield  {author} {\bibinfo {author} {\bibfnamefont {Yu}~\bibnamefont
  {Sun}}, \bibinfo {author} {\bibfnamefont {Zhihao}\ \bibnamefont {Xia}}, \
  and\ \bibinfo {author} {\bibfnamefont {Ulugbek~S}\ \bibnamefont {Kamilov}},\
  }\bibfield  {title} {\enquote {\bibinfo {title} {Efficient and accurate
  inversion of multiple scattering with deep learning},}\ }\href@noop {}
  {\bibfield  {journal} {\bibinfo  {journal} {Optics Express}\ }\textbf
  {\bibinfo {volume} {26}},\ \bibinfo {pages} {14678--14688} (\bibinfo {year}
  {2018})}\BibitemShut {NoStop}%
\bibitem [{\citenamefont {Sanghvi}\ \emph {et~al.}(2019)\citenamefont
  {Sanghvi}, \citenamefont {Kalepu},\ and\ \citenamefont
  {K.~Khankhoje}}]{sanghvi2019embedding}%
  \BibitemOpen
  \bibfield  {author} {\bibinfo {author} {\bibfnamefont {Yash}\ \bibnamefont
  {Sanghvi}}, \bibinfo {author} {\bibfnamefont {Yaswanth}\ \bibnamefont
  {Kalepu}}, \ and\ \bibinfo {author} {\bibfnamefont {Uday}\ \bibnamefont
  {K.~Khankhoje}},\ }\bibfield  {title} {\enquote {\bibinfo {title} {Embedding
  deep learning in inverse scattering problems},}\ }\href@noop {} {\bibfield
  {journal} {\bibinfo  {journal} {IEEE Transactions on Computational Imaging}\
  } (\bibinfo {year} {2019})}\BibitemShut {NoStop}%
\bibitem [{\citenamefont {Raissi}\ \emph {et~al.}(2019)\citenamefont {Raissi},
  \citenamefont {Perdikaris},\ and\ \citenamefont {Karniadakis}}]{Raissi}%
  \BibitemOpen
  \bibfield  {author} {\bibinfo {author} {\bibfnamefont {M.}~\bibnamefont
  {Raissi}}, \bibinfo {author} {\bibfnamefont {P.}~\bibnamefont {Perdikaris}},
  \ and\ \bibinfo {author} {\bibfnamefont {G.~E.}\ \bibnamefont
  {Karniadakis}},\ }\bibfield  {title} {\enquote {\bibinfo {title}
  {Physics-informed neural networks: A deep learning framework for solving
  forward and inverse problems involving nonlinear partial differential
  equations},}\ }\href@noop {} {\bibfield  {journal} {\bibinfo  {journal} {J.
  of Comp. Phys.}\ }\textbf {\bibinfo {volume} {378}},\ \bibinfo {pages}
  {686--707} (\bibinfo {year} {2019})}\BibitemShut {NoStop}%
\bibitem [{\citenamefont {Lu}\ \emph {et~al.}(2019)\citenamefont {Lu},
  \citenamefont {Meng}, \citenamefont {Mao},\ and\ \citenamefont
  {Karniadakis}}]{lu2019deepxde}%
  \BibitemOpen
  \bibfield  {author} {\bibinfo {author} {\bibfnamefont {Lu}~\bibnamefont
  {Lu}}, \bibinfo {author} {\bibfnamefont {Xuhui}\ \bibnamefont {Meng}},
  \bibinfo {author} {\bibfnamefont {Zhiping}\ \bibnamefont {Mao}}, \ and\
  \bibinfo {author} {\bibfnamefont {George~E}\ \bibnamefont {Karniadakis}},\
  }\bibfield  {title} {\enquote {\bibinfo {title} {Deepxde: A deep learning
  library for solving differential equations},}\ }\href@noop {} {\bibfield
  {journal} {\bibinfo  {journal} {arXiv preprint arXiv:1907.04502}\ } (\bibinfo
  {year} {2019})}\BibitemShut {NoStop}%
\bibitem [{\citenamefont {Abadi}\ \emph {et~al.}(2016)\citenamefont {Abadi},
  \citenamefont {Barham}, \citenamefont {Chen}, \citenamefont {Davis},
  \citenamefont {Dean}, \citenamefont {Ghemawat}, \citenamefont {Irving},\ and\
  \citenamefont {Isard}}]{Abadi}%
  \BibitemOpen
  \bibfield  {author} {\bibinfo {author} {\bibfnamefont {M.}~\bibnamefont
  {Abadi}}, \bibinfo {author} {\bibfnamefont {P.}~\bibnamefont {Barham}},
  \bibinfo {author} {\bibfnamefont {Z.}~\bibnamefont {Chen}}, \bibinfo {author}
  {\bibfnamefont {A.}~\bibnamefont {Davis}}, \bibinfo {author} {\bibfnamefont
  {J.}~\bibnamefont {Dean}}, \bibinfo {author} {\bibfnamefont {S.}~\bibnamefont
  {Ghemawat}}, \bibinfo {author} {\bibfnamefont {G.}~\bibnamefont {Irving}}, \
  and\ \bibinfo {author} {\bibfnamefont {M.}~\bibnamefont {Isard}},\ }\bibfield
   {title} {\enquote {\bibinfo {title} {Tensorflow:a system for large-scale
  machine learning},}\ }\href@noop {} {\bibfield  {journal} {\bibinfo
  {journal} {12th USENIX Symposiumon Operating Systems Design and
  Implementation}\ ,\ \bibinfo {pages} {265--283}} (\bibinfo {year}
  {2016})}\BibitemShut {NoStop}%
\bibitem [{\citenamefont {Soukoulis}\ and\ \citenamefont
  {Wegener}(2011)}]{soukoulis2011past}%
  \BibitemOpen
  \bibfield  {author} {\bibinfo {author} {\bibfnamefont {Costas~M}\
  \bibnamefont {Soukoulis}}\ and\ \bibinfo {author} {\bibfnamefont {Martin}\
  \bibnamefont {Wegener}},\ }\bibfield  {title} {\enquote {\bibinfo {title}
  {Past achievements and future challenges in the development of
  three-dimensional photonic metamaterials},}\ }\href@noop {} {\bibfield
  {journal} {\bibinfo  {journal} {Nature Photonics}\ }\textbf {\bibinfo
  {volume} {5}},\ \bibinfo {pages} {523} (\bibinfo {year} {2011})}\BibitemShut
  {NoStop}%
\bibitem [{\citenamefont {Sheng}(2006)}]{sheng2006introduction}%
  \BibitemOpen
  \bibfield  {author} {\bibinfo {author} {\bibfnamefont {Ping}\ \bibnamefont
  {Sheng}},\ }\href@noop {} {\emph {\bibinfo {title} {Introduction to wave
  scattering, localization and mesoscopic phenomena}}},\ Vol.~\bibinfo {volume}
  {88}\ (\bibinfo  {publisher} {Springer Science \& Business Media},\ \bibinfo
  {year} {2006})\BibitemShut {NoStop}%
\bibitem [{\citenamefont {Choy}(2015)}]{choy2015effective}%
  \BibitemOpen
  \bibfield  {author} {\bibinfo {author} {\bibfnamefont {Tuck~C}\ \bibnamefont
  {Choy}},\ }\href@noop {} {\emph {\bibinfo {title} {Effective medium theory:
  principles and applications}}},\ Vol.\ \bibinfo {volume} {165}\ (\bibinfo
  {publisher} {Oxford University},\ \bibinfo {year} {2015})\BibitemShut
  {NoStop}%
\bibitem [{\citenamefont {Sihvola}(1999)}]{sihvola1999electromagnetic}%
  \BibitemOpen
  \bibfield  {author} {\bibinfo {author} {\bibfnamefont {Ari~H}\ \bibnamefont
  {Sihvola}},\ }\href@noop {} {\emph {\bibinfo {title} {Electromagnetic mixing
  formulas and applications}}},\ \bibinfo {number} {47}\ (\bibinfo  {publisher}
  {The Institution of Engineering and Technology},\ \bibinfo {year}
  {1999})\BibitemShut {NoStop}%
\bibitem [{\citenamefont {Wu}\ \emph {et~al.}(2006)\citenamefont {Wu},
  \citenamefont {Li}, \citenamefont {Zhang},\ and\ \citenamefont
  {Chan}}]{wu2006effective}%
  \BibitemOpen
  \bibfield  {author} {\bibinfo {author} {\bibfnamefont {Ying}\ \bibnamefont
  {Wu}}, \bibinfo {author} {\bibfnamefont {Jensen}\ \bibnamefont {Li}},
  \bibinfo {author} {\bibfnamefont {Zhao-Qing}\ \bibnamefont {Zhang}}, \ and\
  \bibinfo {author} {\bibfnamefont {CT}~\bibnamefont {Chan}},\ }\bibfield
  {title} {\enquote {\bibinfo {title} {Effective medium theory for
  magnetodielectric composites: Beyond the long-wavelength limit},}\
  }\href@noop {} {\bibfield  {journal} {\bibinfo  {journal} {Physical Review
  B}\ }\textbf {\bibinfo {volume} {74}},\ \bibinfo {pages} {085111} (\bibinfo
  {year} {2006})}\BibitemShut {NoStop}%
\bibitem [{\citenamefont {Zhang}\ \emph {et~al.}(2013)\citenamefont {Zhang},
  \citenamefont {Shen}, \citenamefont {Xu}, \citenamefont {Zhu}, \citenamefont
  {Lei}, \citenamefont {Zhang},\ and\ \citenamefont {Xu}}]{zhang2013effective}%
  \BibitemOpen
  \bibfield  {author} {\bibinfo {author} {\bibfnamefont {Hao}\ \bibnamefont
  {Zhang}}, \bibinfo {author} {\bibfnamefont {Yongqiang}\ \bibnamefont {Shen}},
  \bibinfo {author} {\bibfnamefont {Yuchen}\ \bibnamefont {Xu}}, \bibinfo
  {author} {\bibfnamefont {Heyuan}\ \bibnamefont {Zhu}}, \bibinfo {author}
  {\bibfnamefont {Ming}\ \bibnamefont {Lei}}, \bibinfo {author} {\bibfnamefont
  {Xiangchao}\ \bibnamefont {Zhang}}, \ and\ \bibinfo {author} {\bibfnamefont
  {Min}\ \bibnamefont {Xu}},\ }\bibfield  {title} {\enquote {\bibinfo {title}
  {Effective medium theory for two-dimensional random media composed of
  core-shell cylinders},}\ }\href@noop {} {\bibfield  {journal} {\bibinfo
  {journal} {Optics Communications}\ }\textbf {\bibinfo {volume} {306}},\
  \bibinfo {pages} {9--16} (\bibinfo {year} {2013})}\BibitemShut {NoStop}%
\bibitem [{\citenamefont {Chremmos}\ \emph {et~al.}(2015)\citenamefont
  {Chremmos}, \citenamefont {Kallos}, \citenamefont {Giamalaki}, \citenamefont
  {Yannopapas},\ and\ \citenamefont {Paspalakis}}]{chremmos2015effective}%
  \BibitemOpen
  \bibfield  {author} {\bibinfo {author} {\bibfnamefont {Ioannis}\ \bibnamefont
  {Chremmos}}, \bibinfo {author} {\bibfnamefont {Efthymios}\ \bibnamefont
  {Kallos}}, \bibinfo {author} {\bibfnamefont {Melpomeni}\ \bibnamefont
  {Giamalaki}}, \bibinfo {author} {\bibfnamefont {Vassilios}\ \bibnamefont
  {Yannopapas}}, \ and\ \bibinfo {author} {\bibfnamefont {Emmanuel}\
  \bibnamefont {Paspalakis}},\ }\bibfield  {title} {\enquote {\bibinfo {title}
  {Effective medium theory for two-dimensional non-magnetic metamaterial
  lattices up to quadrupole expansions},}\ }\href@noop {} {\bibfield  {journal}
  {\bibinfo  {journal} {Journal of Optics}\ }\textbf {\bibinfo {volume} {17}},\
  \bibinfo {pages} {075102} (\bibinfo {year} {2015})}\BibitemShut {NoStop}%
\bibitem [{\citenamefont {Gozhenko}\ \emph {et~al.}(2013)\citenamefont
  {Gozhenko}, \citenamefont {Amert},\ and\ \citenamefont
  {Whites}}]{gozhenko2013homogenization}%
  \BibitemOpen
  \bibfield  {author} {\bibinfo {author} {\bibfnamefont {Victor~V}\
  \bibnamefont {Gozhenko}}, \bibinfo {author} {\bibfnamefont {Anthony~K}\
  \bibnamefont {Amert}}, \ and\ \bibinfo {author} {\bibfnamefont {Keith~W}\
  \bibnamefont {Whites}},\ }\bibfield  {title} {\enquote {\bibinfo {title}
  {Homogenization of periodic metamaterials by field averaging over unit cell
  boundaries: use and limitations},}\ }\href@noop {} {\bibfield  {journal}
  {\bibinfo  {journal} {New Journal of Physics}\ }\textbf {\bibinfo {volume}
  {15}},\ \bibinfo {pages} {043030} (\bibinfo {year} {2013})}\BibitemShut
  {NoStop}%
\bibitem [{\citenamefont {Trevino}\ \emph
  {et~al.}(2012{\natexlab{a}})\citenamefont {Trevino}, \citenamefont {Liew},
  \citenamefont {Noh}, \citenamefont {Cao},\ and\ \citenamefont
  {Dal~Negro}}]{trevino2012geometrical}%
  \BibitemOpen
  \bibfield  {author} {\bibinfo {author} {\bibfnamefont {Jacob}\ \bibnamefont
  {Trevino}}, \bibinfo {author} {\bibfnamefont {Seng~Fatt}\ \bibnamefont
  {Liew}}, \bibinfo {author} {\bibfnamefont {Heeso}\ \bibnamefont {Noh}},
  \bibinfo {author} {\bibfnamefont {Hui}\ \bibnamefont {Cao}}, \ and\ \bibinfo
  {author} {\bibfnamefont {Luca}\ \bibnamefont {Dal~Negro}},\ }\bibfield
  {title} {\enquote {\bibinfo {title} {Geometrical structure, multifractal
  spectra and localized optical modes of aperiodic vogel spirals},}\
  }\href@noop {} {\bibfield  {journal} {\bibinfo  {journal} {Optics Express}\
  }\textbf {\bibinfo {volume} {20}},\ \bibinfo {pages} {3015--3033} (\bibinfo
  {year} {2012}{\natexlab{a}})}\BibitemShut {NoStop}%
\bibitem [{\citenamefont {Lawrence}\ \emph {et~al.}(2012)\citenamefont
  {Lawrence}, \citenamefont {Trevino},\ and\ \citenamefont
  {Dal~Negro}}]{lawrence2012control}%
  \BibitemOpen
  \bibfield  {author} {\bibinfo {author} {\bibfnamefont {Nate}\ \bibnamefont
  {Lawrence}}, \bibinfo {author} {\bibfnamefont {Jacob}\ \bibnamefont
  {Trevino}}, \ and\ \bibinfo {author} {\bibfnamefont {Luca}\ \bibnamefont
  {Dal~Negro}},\ }\bibfield  {title} {\enquote {\bibinfo {title} {Control of
  optical orbital angular momentum by vogel spiral arrays of metallic
  nanoparticles},}\ }\href@noop {} {\bibfield  {journal} {\bibinfo  {journal}
  {Optics Letters}\ }\textbf {\bibinfo {volume} {37}},\ \bibinfo {pages}
  {5076--5078} (\bibinfo {year} {2012})}\BibitemShut {NoStop}%
\bibitem [{\citenamefont {Pollard}\ and\ \citenamefont
  {Parker}(2009)}]{pollard2009low}%
  \BibitemOpen
  \bibfield  {author} {\bibinfo {author} {\bibfnamefont {Michael~E}\
  \bibnamefont {Pollard}}\ and\ \bibinfo {author} {\bibfnamefont {Gregory~J}\
  \bibnamefont {Parker}},\ }\bibfield  {title} {\enquote {\bibinfo {title}
  {Low-contrast bandgaps of a planar parabolic spiral lattice},}\ }\href@noop
  {} {\bibfield  {journal} {\bibinfo  {journal} {Optics Letters}\ }\textbf
  {\bibinfo {volume} {34}},\ \bibinfo {pages} {2805--2807} (\bibinfo {year}
  {2009})}\BibitemShut {NoStop}%
\bibitem [{\citenamefont {Trevino}\ \emph {et~al.}(2011)\citenamefont
  {Trevino}, \citenamefont {Cao},\ and\ \citenamefont
  {Dal~Negro}}]{trevino2011circularly}%
  \BibitemOpen
  \bibfield  {author} {\bibinfo {author} {\bibfnamefont {Jacob}\ \bibnamefont
  {Trevino}}, \bibinfo {author} {\bibfnamefont {Hui}\ \bibnamefont {Cao}}, \
  and\ \bibinfo {author} {\bibfnamefont {Luca}\ \bibnamefont {Dal~Negro}},\
  }\bibfield  {title} {\enquote {\bibinfo {title} {Circularly symmetric light
  scattering from nanoplasmonic spirals},}\ }\href@noop {} {\bibfield
  {journal} {\bibinfo  {journal} {Nano Letters}\ }\textbf {\bibinfo {volume}
  {11}},\ \bibinfo {pages} {2008--2016} (\bibinfo {year} {2011})}\BibitemShut
  {NoStop}%
\bibitem [{\citenamefont {Liew}\ \emph {et~al.}(2011)\citenamefont {Liew},
  \citenamefont {Noh}, \citenamefont {Trevino}, \citenamefont {Dal~Negro},\
  and\ \citenamefont {Cao}}]{liew2011localized}%
  \BibitemOpen
  \bibfield  {author} {\bibinfo {author} {\bibfnamefont {Seng~Fatt}\
  \bibnamefont {Liew}}, \bibinfo {author} {\bibfnamefont {Heeso}\ \bibnamefont
  {Noh}}, \bibinfo {author} {\bibfnamefont {Jacob}\ \bibnamefont {Trevino}},
  \bibinfo {author} {\bibfnamefont {Luca}\ \bibnamefont {Dal~Negro}}, \ and\
  \bibinfo {author} {\bibfnamefont {Hui}\ \bibnamefont {Cao}},\ }\bibfield
  {title} {\enquote {\bibinfo {title} {Localized photonic band edge modes and
  orbital angular momenta of light in a golden-angle spiral},}\ }\href@noop {}
  {\bibfield  {journal} {\bibinfo  {journal} {Optics Express}\ }\textbf
  {\bibinfo {volume} {19}},\ \bibinfo {pages} {23631--23642} (\bibinfo {year}
  {2011})}\BibitemShut {NoStop}%
\bibitem [{\citenamefont {Trevino}\ \emph
  {et~al.}(2012{\natexlab{b}})\citenamefont {Trevino}, \citenamefont
  {Forestiere}, \citenamefont {Di~Martino}, \citenamefont {Yerci},
  \citenamefont {Priolo},\ and\ \citenamefont
  {Dal~Negro}}]{trevino2012plasmonic}%
  \BibitemOpen
  \bibfield  {author} {\bibinfo {author} {\bibfnamefont {Jacob}\ \bibnamefont
  {Trevino}}, \bibinfo {author} {\bibfnamefont {Carlo}\ \bibnamefont
  {Forestiere}}, \bibinfo {author} {\bibfnamefont {Giuliana}\ \bibnamefont
  {Di~Martino}}, \bibinfo {author} {\bibfnamefont {Selcuk}\ \bibnamefont
  {Yerci}}, \bibinfo {author} {\bibfnamefont {Francesco}\ \bibnamefont
  {Priolo}}, \ and\ \bibinfo {author} {\bibfnamefont {Luca}\ \bibnamefont
  {Dal~Negro}},\ }\bibfield  {title} {\enquote {\bibinfo {title}
  {Plasmonic-photonic arrays with aperiodic spiral order for ultra-thin film
  solar cells},}\ }\href@noop {} {\bibfield  {journal} {\bibinfo  {journal}
  {Optics Express}\ }\textbf {\bibinfo {volume} {20}},\ \bibinfo {pages}
  {A418--A430} (\bibinfo {year} {2012}{\natexlab{b}})}\BibitemShut {NoStop}%
\bibitem [{\citenamefont {Razi}\ \emph {et~al.}(2019)\citenamefont {Razi},
  \citenamefont {Wang}, \citenamefont {He}, \citenamefont {Kirby},\ and\
  \citenamefont {Dal~Negro}}]{razi2019optimization}%
  \BibitemOpen
  \bibfield  {author} {\bibinfo {author} {\bibfnamefont {Mani}\ \bibnamefont
  {Razi}}, \bibinfo {author} {\bibfnamefont {Ren}\ \bibnamefont {Wang}},
  \bibinfo {author} {\bibfnamefont {Yanyan}\ \bibnamefont {He}}, \bibinfo
  {author} {\bibfnamefont {Robert~M}\ \bibnamefont {Kirby}}, \ and\ \bibinfo
  {author} {\bibfnamefont {Luca}\ \bibnamefont {Dal~Negro}},\ }\bibfield
  {title} {\enquote {\bibinfo {title} {Optimization of large-scale vogel spiral
  arrays of plasmonic nanoparticles},}\ }\href@noop {} {\bibfield  {journal}
  {\bibinfo  {journal} {Plasmonics}\ }\textbf {\bibinfo {volume} {14}},\
  \bibinfo {pages} {253--261} (\bibinfo {year} {2019})}\BibitemShut {NoStop}%
\bibitem [{\citenamefont {Sgrignuoli}\ \emph {et~al.}(2019)\citenamefont
  {Sgrignuoli}, \citenamefont {Wang}, \citenamefont {Pinheiro},\ and\
  \citenamefont {Dal~Negro}}]{fab2019localization}%
  \BibitemOpen
  \bibfield  {author} {\bibinfo {author} {\bibfnamefont {F.}~\bibnamefont
  {Sgrignuoli}}, \bibinfo {author} {\bibfnamefont {R.}~\bibnamefont {Wang}},
  \bibinfo {author} {\bibfnamefont {F.~A.}\ \bibnamefont {Pinheiro}}, \ and\
  \bibinfo {author} {\bibfnamefont {L.}~\bibnamefont {Dal~Negro}},\ }\bibfield
  {title} {\enquote {\bibinfo {title} {Localization of scattering resonances in
  aperiodic vogel spirals},}\ }\href {\doibase 10.1103/PhysRevB.99.104202}
  {\bibfield  {journal} {\bibinfo  {journal} {Phys. Rev. B}\ }\textbf {\bibinfo
  {volume} {99}},\ \bibinfo {pages} {104202} (\bibinfo {year}
  {2019})}\BibitemShut {NoStop}%
\bibitem [{\citenamefont {Adam}(2011)}]{adam2011mathematical}%
  \BibitemOpen
  \bibfield  {author} {\bibinfo {author} {\bibfnamefont {John~A}\ \bibnamefont
  {Adam}},\ }\href@noop {} {\emph {\bibinfo {title} {A mathematical nature
  walk}}}\ (\bibinfo  {publisher} {Princeton University},\ \bibinfo {year}
  {2011})\BibitemShut {NoStop}%
\bibitem [{\citenamefont {Rechtsman}\ and\ \citenamefont
  {Torquato}(2008)}]{rechtsman_effective_2008}%
  \BibitemOpen
  \bibfield  {author} {\bibinfo {author} {\bibfnamefont {Mikael~C.}\
  \bibnamefont {Rechtsman}}\ and\ \bibinfo {author} {\bibfnamefont {Salvatore}\
  \bibnamefont {Torquato}},\ }\bibfield  {title} {\enquote {\bibinfo {title}
  {Effective dielectric tensor for electromagnetic wave propagation in random
  media},}\ }\href {\doibase 10.1063/1.2906135} {\bibfield  {journal} {\bibinfo
   {journal} {Journal of Applied Physics}\ }\textbf {\bibinfo {volume} {103}},\
  \bibinfo {pages} {084901} (\bibinfo {year} {2008})}\BibitemShut {NoStop}%
\bibitem [{\citenamefont {Kim}\ and\ \citenamefont
  {Torquato}(2019)}]{kim_multifunctional_2019}%
  \BibitemOpen
  \bibfield  {author} {\bibinfo {author} {\bibfnamefont {Jaeuk}\ \bibnamefont
  {Kim}}\ and\ \bibinfo {author} {\bibfnamefont {Salvatore}\ \bibnamefont
  {Torquato}},\ }\bibfield  {title} {\enquote {\bibinfo {title}
  {Multifunctional {{Composites}} for {{Elastic}} and {{Electromagnetic Wave
  Propagation}}},}\ }\href@noop {} {\bibfield  {journal} {\bibinfo  {journal}
  {arXiv:1908.06662}\ } (\bibinfo {year} {2019})},\ \Eprint
  {http://arxiv.org/abs/1908.06662} {arXiv:1908.06662} \BibitemShut {NoStop}%
\bibitem [{\citenamefont {Novotny}\ and\ \citenamefont
  {Hecht}(2012)}]{novotny2012principles}%
  \BibitemOpen
  \bibfield  {author} {\bibinfo {author} {\bibfnamefont {Lukas}\ \bibnamefont
  {Novotny}}\ and\ \bibinfo {author} {\bibfnamefont {Bert}\ \bibnamefont
  {Hecht}},\ }\href@noop {} {\emph {\bibinfo {title} {Principles of
  nano-optics}}}\ (\bibinfo  {publisher} {Cambridge University},\ \bibinfo
  {year} {2012})\BibitemShut {NoStop}%
\bibitem [{\citenamefont {Bohren}\ and\ \citenamefont
  {Huffman}(2008)}]{bohren2008absorption}%
  \BibitemOpen
  \bibfield  {author} {\bibinfo {author} {\bibfnamefont {Craig~F}\ \bibnamefont
  {Bohren}}\ and\ \bibinfo {author} {\bibfnamefont {Donald~R}\ \bibnamefont
  {Huffman}},\ }\href@noop {} {\emph {\bibinfo {title} {Absorption and
  scattering of light by small particles}}}\ (\bibinfo  {publisher} {John Wiley
  \& Sons},\ \bibinfo {year} {2008})\BibitemShut {NoStop}%
\bibitem [{\citenamefont {Geffrin}\ \emph {et~al.}(2005)\citenamefont
  {Geffrin}, \citenamefont {Sabouroux},\ and\ \citenamefont
  {Eyraud}}]{geffrin2005free}%
  \BibitemOpen
  \bibfield  {author} {\bibinfo {author} {\bibfnamefont {Jean-Michel}\
  \bibnamefont {Geffrin}}, \bibinfo {author} {\bibfnamefont {Pierre}\
  \bibnamefont {Sabouroux}}, \ and\ \bibinfo {author} {\bibfnamefont
  {Christelle}\ \bibnamefont {Eyraud}},\ }\bibfield  {title} {\enquote
  {\bibinfo {title} {Free space experimental scattering database continuation:
  experimental set-up and measurement precision},}\ }\href@noop {} {\bibfield
  {journal} {\bibinfo  {journal} {Inverse Problems}\ }\textbf {\bibinfo
  {volume} {21}},\ \bibinfo {pages} {S117} (\bibinfo {year}
  {2005})}\BibitemShut {NoStop}%
\bibitem [{\citenamefont {Hecht}\ \emph {et~al.}(2000)\citenamefont {Hecht},
  \citenamefont {Sick}, \citenamefont {Wild}, \citenamefont {Deckert},
  \citenamefont {Zenobi}, \citenamefont {Martin},\ and\ \citenamefont
  {Pohl}}]{hecht2000scanning}%
  \BibitemOpen
  \bibfield  {author} {\bibinfo {author} {\bibfnamefont {Bert}\ \bibnamefont
  {Hecht}}, \bibinfo {author} {\bibfnamefont {Beate}\ \bibnamefont {Sick}},
  \bibinfo {author} {\bibfnamefont {Urs~P}\ \bibnamefont {Wild}}, \bibinfo
  {author} {\bibfnamefont {Volker}\ \bibnamefont {Deckert}}, \bibinfo {author}
  {\bibfnamefont {Renato}\ \bibnamefont {Zenobi}}, \bibinfo {author}
  {\bibfnamefont {Olivier~JF}\ \bibnamefont {Martin}}, \ and\ \bibinfo {author}
  {\bibfnamefont {Dieter~W}\ \bibnamefont {Pohl}},\ }\bibfield  {title}
  {\enquote {\bibinfo {title} {Scanning near-field optical microscopy with
  aperture probes: Fundamentals and applications},}\ }\href@noop {} {\bibfield
  {journal} {\bibinfo  {journal} {The Journal of Chemical Physics}\ }\textbf
  {\bibinfo {volume} {112}},\ \bibinfo {pages} {7761--7774} (\bibinfo {year}
  {2000})}\BibitemShut {NoStop}%
\bibitem [{\citenamefont {Al{\`u}}\ and\ \citenamefont
  {Engheta}(2008)}]{alu2008plasmonic}%
  \BibitemOpen
  \bibfield  {author} {\bibinfo {author} {\bibfnamefont {Andrea}\ \bibnamefont
  {Al{\`u}}}\ and\ \bibinfo {author} {\bibfnamefont {Nader}\ \bibnamefont
  {Engheta}},\ }\bibfield  {title} {\enquote {\bibinfo {title} {Plasmonic and
  metamaterial cloaking: physical mechanisms and potentials},}\ }\href@noop {}
  {\bibfield  {journal} {\bibinfo  {journal} {Journal of Optics A: Pure and
  Applied Optics}\ }\textbf {\bibinfo {volume} {10}},\ \bibinfo {pages}
  {093002} (\bibinfo {year} {2008})}\BibitemShut {NoStop}%
\bibitem [{\citenamefont {Fleury}\ \emph {et~al.}(2015)\citenamefont {Fleury},
  \citenamefont {Monticone},\ and\ \citenamefont
  {Al{\`u}}}]{fleury2015invisibility}%
  \BibitemOpen
  \bibfield  {author} {\bibinfo {author} {\bibfnamefont {Romain}\ \bibnamefont
  {Fleury}}, \bibinfo {author} {\bibfnamefont {Francesco}\ \bibnamefont
  {Monticone}}, \ and\ \bibinfo {author} {\bibfnamefont {Andrea}\ \bibnamefont
  {Al{\`u}}},\ }\bibfield  {title} {\enquote {\bibinfo {title} {Invisibility
  and cloaking: Origins, present, and future perspectives},}\ }\href@noop {}
  {\bibfield  {journal} {\bibinfo  {journal} {Physical Review Applied}\
  }\textbf {\bibinfo {volume} {4}},\ \bibinfo {pages} {037001} (\bibinfo {year}
  {2015})}\BibitemShut {NoStop}%
\bibitem [{\citenamefont {Al{\`u}}\ \emph {et~al.}(2010)\citenamefont
  {Al{\`u}}, \citenamefont {Rainwater},\ and\ \citenamefont
  {Kerkhoff}}]{alu2010plasmonic}%
  \BibitemOpen
  \bibfield  {author} {\bibinfo {author} {\bibfnamefont {Andrea}\ \bibnamefont
  {Al{\`u}}}, \bibinfo {author} {\bibfnamefont {David}\ \bibnamefont
  {Rainwater}}, \ and\ \bibinfo {author} {\bibfnamefont {Aaron}\ \bibnamefont
  {Kerkhoff}},\ }\bibfield  {title} {\enquote {\bibinfo {title} {Plasmonic
  cloaking of cylinders: finite length, oblique illumination and
  cross-polarization coupling},}\ }\href@noop {} {\bibfield  {journal}
  {\bibinfo  {journal} {New Journal of Physics}\ }\textbf {\bibinfo {volume}
  {12}},\ \bibinfo {pages} {103028} (\bibinfo {year} {2010})}\BibitemShut
  {NoStop}%
\bibitem [{\citenamefont {Vanschoren}(2018)}]{vanschoren2018metalearning}%
  \BibitemOpen
  \bibfield  {author} {\bibinfo {author} {\bibfnamefont {Joaquin}\ \bibnamefont
  {Vanschoren}},\ }\bibfield  {title} {\enquote {\bibinfo {title}
  {Meta-learning: A survey},}\ }\href@noop {} {\bibfield  {journal} {\bibinfo
  {journal} {arXiv preprint arXiv:1810.03548}\ } (\bibinfo {year}
  {2018})}\BibitemShut {NoStop}%
\bibitem [{\citenamefont {Chen}\ \emph {et~al.}(2019)\citenamefont {Chen},
  \citenamefont {Duan},\ and\ \citenamefont {Karniadakis}}]{chen2019learning}%
  \BibitemOpen
  \bibfield  {author} {\bibinfo {author} {\bibfnamefont {Xiaoli}\ \bibnamefont
  {Chen}}, \bibinfo {author} {\bibfnamefont {Jinqiao}\ \bibnamefont {Duan}}, \
  and\ \bibinfo {author} {\bibfnamefont {George~Em}\ \bibnamefont
  {Karniadakis}},\ }\bibfield  {title} {\enquote {\bibinfo {title} {Learning
  and meta-learning of stochastic advection-diffusion-reaction systems from
  sparse measurements},}\ }\href@noop {} {\bibfield  {journal} {\bibinfo
  {journal} {arXiv preprint arXiv:1910.09098}\ } (\bibinfo {year}
  {2019})}\BibitemShut {NoStop}%
\bibitem [{\citenamefont {He}\ \emph {et~al.}(2019)\citenamefont {He},
  \citenamefont {Zhao},\ and\ \citenamefont {Chu}}]{he2019automl}%
  \BibitemOpen
  \bibfield  {author} {\bibinfo {author} {\bibfnamefont {Xin}\ \bibnamefont
  {He}}, \bibinfo {author} {\bibfnamefont {Kaiyong}\ \bibnamefont {Zhao}}, \
  and\ \bibinfo {author} {\bibfnamefont {Xiaowen}\ \bibnamefont {Chu}},\
  }\bibfield  {title} {\enquote {\bibinfo {title} {Automl: A survey of the
  state-of-the-art},}\ }\href@noop {} {\bibfield  {journal} {\bibinfo
  {journal} {arXiv preprint arXiv:1908.00709}\ } (\bibinfo {year}
  {2019})}\BibitemShut {NoStop}%
\end{thebibliography}
\end{document}